\documentclass[]{interact}

\usepackage{epstopdf}
\usepackage{subfigure}
\usepackage{eurosym}
\usepackage{siunitx}
\DeclareMathOperator*{\argmin}{arg\,min}

\usepackage{natbib}
\bibpunct[, ]{(}{)}{;}{a}{}{,}
\renewcommand\bibfont{\fontsize{10}{12}\selectfont}
\newcommand*\rot{\rotatebox{90}}

\theoremstyle{plain}

\theoremstyle{definition}

\theoremstyle{remark}

\usepackage{xcolor}

\begin{document}


\title{Evolution of Labour Supply in Ridesourcing}

\author{
\name{Arjan de Ruijter\textsuperscript{a}\thanks{CONTACT Arjan de Ruijter. Email: a.j.f.deruijter@tudelft.nl}, Oded Cats\textsuperscript{a}, Rafał Kucharski\textsuperscript{a} and Hans van Lint\textsuperscript{a}}
\affil{\textsuperscript{a}Delft University of Technology, Delft, Netherlands}
}

\maketitle

\begin{abstract}
Contrary to traditional transit services, supply in ridesourcing systems emerges from individual labour decisions of gig workers. The effect of decentralization in supply on the evolution of on-demand transit services is largely unknown. To this end, we propose a dynamic model comprising of the subsequent supply-side processes: (i) initial exposure to information about the platform, (ii) a long-term registration decision, and (iii) daily participation decisions, subject to day-to-day learning based on within-day matching outcomes. We construct a series of experiments to study the effect of supply market properties and pricing strategies. We find that labour supply in ridesourcing may be non-linear and undergo several transitions, inducing significant variations in income levels and level of service over time. Our results provide indications that the ridesourcing market may benefit from a cap in supply and regulation of the commission fee.
\end{abstract}


\section{Introduction}
In many service industries the role of businesses is shifting from service provision to facilitating the exchange of services, typically through the creation of virtual two-sided marketplaces. When suppliers in a two-sided market are individual contractors rather than businesses, the market is considered to be part of the gig economy. In contrast to traditional fixed labour contracts offering long-term security to all parties involved, labour in the gig economy is organised through more flexible arrangements. Not only does this allow service providers to respond more adequately to changes in demand than operators with more traditional forms of labour, it also means that they may be exempted from paying employee benefits \citep{Pra15}. Unfortunately, recent protests demonstrate that the value gig workers derive from the flexibility to set their own working schedules \citep{hall2018analysis,Chen2019value} may not outweigh the loss of financial security associated with flexible labour agreements. While the social desirability of these new forms of labour agreements is disputed, the gig economy has gained ground across many industries.

Transportation is a predominant example with platform businesses in food delivery (Just Eat Takeaway, Uber Eats, DoorDash), package delivery (Amazon Flex) and passenger services (Uber, Lyft, DiDi). Service providers in the third category are commonly referred to as ridesourcing providers or Transport Network Companies (TNCs). Typically, ridesourcing businesses reward drivers based on satisfied demand rather than based on time spent working for the platform. Hence, in contrast to traditional transit operators with employed drivers, they do not bear the cost of excess labour available through their platform. This is beneficial especially in times of rapidly declining travel demand, such as during the COVID-19 pandemic. 

The question remains whether a decentralisation of supply is truly a win-win for service provider, suppliers and consumers in the ride-hailing market. Early evidence suggests that in addition to losing access to social provisions related to employment, ridesourcing drivers may receive inadequate financial compensation for supplied labour. In Chicago for example, strong competition between suppliers has led to average driver earnings below the local minimum wage \citep{henao2019analysis}. Besides suppressing driver earnings, oversupply contributes to road congestion by inducing repositioning by idle drivers waiting to be matched \citep{beojone2021inefficiency}. Travellers on the other hand may benefit from an oversupplied market through low waiting times and few denied requests.

Sustained supply of labour to a platform with low payouts suggests that the tragedy of the commons may apply to the ridesourcing market. It occurs when excessive participation leads to a depletion of the total value derived from participation on the platform. A potential reason why drivers may continue to participate under these conditions is that they have limited alternative opportunities in the labour market. Oversupply in the ridesourcing market may also be explained by large temporal variations in labour opportunity costs underlying the value of flexible work \citep{Chen2019value,ashkrof2020understanding}. When a potential driver is not involved in alternative activities - such as alternative employment or education - on a particular day, (s)he may be tempted to work for the platform even when expected earnings are low. In other words, varying opportunity costs caused by activity schedules may disturb the balancing loop of competition in labour supply. 

In contrast, ridesourcing platforms may struggle to attract enough suppliers to the market when the labour market is strong, especially when employment yields high social security benefits. This hampers travellers' chances to find a (quick and cheap) ride. When ride requests have to be rejected or when travellers stop making requests altogether, the service provider is confronted with lost revenue. This may ultimately result in the termination of the service. \cite{farrell2017online} have observed that the growth of on-demand service platforms in many cities is indeed limited by the availability of workers rather than customers. 

In order to comprehend the societal implications of ridesourcing, we thus need to understand how the decentralisation of supply affects the fleet size of a ride-hailing service. Considering the bottom-up nature of ridesourcing supply, its analysis requires investigating system-level effects of factors influencing individual driver's labour decisions. This includes not only strategical decisions by the platform, but also labour market properties and driver characteristics. In this study, we therefore focus on structural supply deficits / surpluses that may exist in the ridesourcing market. Hence, we study labour supply only at the extensive margin as opposed to highly temporal imbalances in supply and demand which may follow from hourly variations in opportunity costs and/or travel demand, i.e. we will capture how many drivers work on a day, but not how long they work on that day.

\subsection{Ridesourcing system dynamics}
The emergence of ridesourcing has not gone unnoticed in the scientific community. In a review of ridesourcing literature, \cite{wang2019ridesourcing} have identified four major research problems related to the impact of emerging ridesourcing services. These topics include the effect of ridesourcing on other modes of transportation \citep{qian2017taxi,zhu2020analysis,ke2020ride,yu2020balancing,ke2021equilibrium}, its broader societal and environmental impacts \citep{rayle2016just,clewlow2017disruptive,yu2017environmental,jin2018ridesourcing}, competition between service providers \citep{zha2016economic,zhou2020competitive}, and the effectiveness of regulations in the ridesourcing market \citep{zha2018surge,zha2018geometric,li2019regulating,yu2020balancing}. A key factor when identifying the societal impacts of ridesourcing is the pricing strategy adopted by the service provider. Hence, many studies revolve around the optimisation of ridesourcing pricing strategies, including the specification of ride fares and driver wages  \citep{banerjee2015pricing,taylor2018demand,zha2018surge,zha2018geometric,bai2019coordinating,sun2019optimal,bimpikis2019spatial,nourinejad2020ride,dong2021optimal}. 

A common feature of previously mentioned works is that the ridesourcing market is represented using a static steady-state model. While allowing insightful analyses into ridesourcing equilibria, there are two downsides to this approach. First, static models are incapable of explaining system evolution towards proposed equilibria. Second, these models fail to capture key dynamic processes that are inherent to ridesourcing provision. Arguably, the equilibria sketched in previous studies may not be realised in practice. In the following, we distinguish several complex day-to-day processes underlying the emergence of decentralised ridesourcing supply.

First, labour supply decisions are affected by a driver's participation history. Because there is no guaranteed participation reward and drivers lack proper means of communicating with other drivers \citep{robinson2017making}, drivers' own experiences form an important source of information in the participation decision. Given that ridesourcing earnings are highly sensitive to system variables such as travel demand and other drivers' labour decisions \citep{bokanyi2020understanding}, there may be large day-to-day variations in the average participation reward. Moreover, due to path-dependent spatial relations between successive matchings of drivers and travel requests, ridesourcing earnings may be distributed unevenly among participating drivers. To illustrate, a driver who is assigned to deliver a passenger in a low demand area may struggle to find a subsequent ride. 'Unlucky' individuals with below average earnings may decide to leave the platform before learning that the system average earnings were higher than their personal earnings. Hence, the unpredictability of ridesourcing earnings can affect the amount of labour available for platform operations.

Second, participation may require making financial investments or entering into contracts. Even though entry barriers for ridesourcing are typically lower than those for conventional taxis \citep{hall2018analysis}, empirical findings still show an increase in vehicle ownership in the population associated with the launch of a ridesourcing service \citep{gong2017uber}. This demonstrates that ridesourcing drivers do not necessarily drive for the platform with a vehicle they already owned. In addition, a taxi license or appropriate driver insurance may need to be obtained to enter the ridesourcing market \citep{baron2018disruptive}. Hence, participation decisions are preceded by a registration decision in which required investments are traded off against anticipated future revenues from participation. The discrepancy in costs between registration (with entry costs) and participation (when entry costs are sunk) implies that studies neglecting registration choice may either overestimate or underestimate the ridesourcing fleet size. This depends on whether the drop in the number of registered drivers outweighs more frequent participation by registered drivers to compensate for the capital costs associated with platform registration \citep{hall2018analysis}. 

Based on a theory of innovation diffusion \citep{rogers2010diffusion}, there are two more steps preceding drivers' platform participation choice: (1) becoming aware of its existence and (2) being persuaded to gather more information about its utility. Variations in attitudes, preferences and social network may explain why individual agents may undergo these stages at different moments in time. The rate at which potential drivers may start considering registration is relevant because a very rapid increase in supply may lead to sharply decreasing participation earnings. A slow diffusion on the other hand may lead to a prolonged situation with long waiting times and therefore dissatisfied travellers. 

To gain a better understanding of equilibria in ridesourcing systems, we need dynamic models that can account for the previously mentioned processes in drivers' labour supply decisions. To the best of our knowledge, applications of day-to-day learning models for ridesourcing systems have been limited to only a few studies. 

One of these has represented ridesourcing evolution with learning behaviour by drivers. \citet{djavadian2017agent} proposed a stochastic day-to-day approach with an integrated within-day operating policy, in which travellers choose ridesourcing if it maximises their expected consumer surplus, anticipating travel time based on experience. Drivers supply labour when their learned perceived income exceeds a deterministic income threshold, implying that variables other than expected income that play a role in drivers' labour supply decisions are neglected. The model proposed by \cite{djavadian2017agent} also does not account for the stages preceding participation, such as the registration process. The method is applied only to a minimal case study representing access to and egress from a single railway station, with supply levels limited to 20 drivers or lower. 

\cite{cachon2017role} and \cite{yu2020balancing} propose a semi-dynamic model consisting of a registration phase and a participation phase. Both phases are strictly separated in time, which means that the model cannot capture interactions between participation decisions of existing drivers and the registration decisions of potential drivers. \cite{dong2021optimal} apply a similar methodology when studying ridesourcing service providers opting for a dual-sourcing strategy. Drivers in their study first decide whether they take up an employment offer by the provider, giving up work schedule flexibility in return for reduced income uncertainty. In the second phase, those that rejected the offer decide on platform participation. Drivers are only offered employment once, i.e. there is no feedback loop between participation and employment. The aforementioned studies apply macroscopic models to represent the within-day matching process, neglecting complex disaggregate spatio-temporal within-day relations between supply, demand and service provider that influence drivers' labour supply decisions.

\subsection{Study contributions}
We represent the long-term evolution of ridesourcing supply by explicitly considering complex interactions between within-day ride-hailing operations, registration barriers and day-to-day variations in opportunity costs. We do so by proposing a day-to-day learning model with a decentralised labour supply, explicitly distinguishing between two dimensions: registration and participation. For platform registration, we develop a probabilistic agent-based model that accounts for registration costs, opportunity costs and anticipated income levels. We propose a macroscopic model based on an epidemiological process to represent diffusion of information between registered and non-registered drivers, concerning the awareness of and satisfaction with the ridesourcing platform. For the daily participation decision, we establish a probabilistic agent-based choice model that acknowledges that drivers' daily participation decision is not merely based on the expected income on a given day derived from accumulated day-to-day experience, but which also depends on unobserved factors such as variations in opportunity costs. 

We integrate our day-to-day model into MaaSSim, an agent-based discrete event simulator of mobility-on-demand operations \citep{Kucharski2020MaaSSim}. The agent-based nature of this model allows us to capture heterogeneity in ridesourcing earnings following from disaggregate and spatially-dependent interactions between demand, supply and platform dynamics in ridesourcing operations, which may affect the emergent ridesourcing fleet.

The model is applied to a case study representing a realistic urban network, with up to 1000 vehicles, to allow for the examination of emergence properties in a decentralised supply market in ridesourcing provision. More specifically, we construct an experiment to find the extent to which labour supply in the market is dependent on the availability and cost of labour in the market. This allows us to answer whether ridesourcing provision risks attaining undesired levels of supply, i.e. over- or under-supplied. In addition, our experiment includes an investigation of the commission rate charged by the platform, in order to explore the implications of profit maximisation in a decentralised market, for both drivers and travellers. Other variables that we study are platform registration barriers and variability in drivers' daily opportunity costs, in order to understand how they characterise supply in ridesourcing provision. Finally, we employ an exhaustive search for establishing the optimal ridesourcing fleet size for travellers, drivers and service provider, which we compare to the equilibrium participation volume in decentralised ridesourcing provision.

\section{Methodology}
\begin{figure}[!ht]
  \centering
  \includegraphics[width=0.85\textwidth]{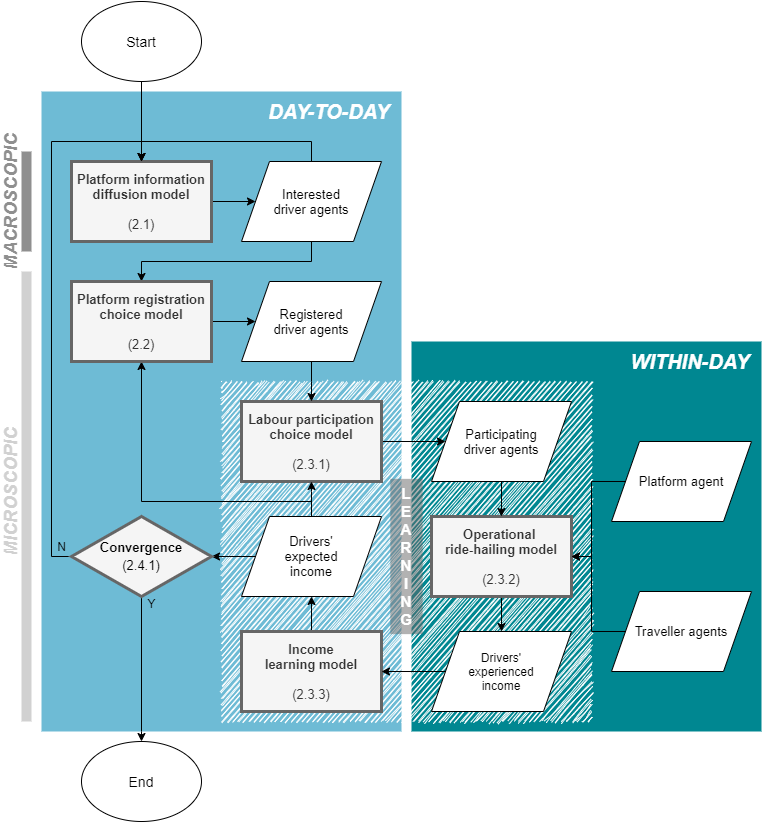}
  \caption{Conceptual framework of the proposed dynamic ridesourcing model, including references to subsections in which a particular submodel is explained}\label{fig:Conc-fw}
\end{figure}

We develop an agent-based day-to-day model with driver agents potentially willing to work for the platform. These agents are at any given moment in time in one of three states: uninformed, interested or registered. Uninformed driver agents are potential drivers currently unaware of the existence of the service. Interested drivers are those that have been informed about the existence of the platform and now monitor the average participation reward. They make an occasional registration decision. Once drivers are registered, they make a daily participation choice, based on the expected income that is learned from previous driving experiences. This is simulated by integrating our day-to-day model, comprising of information diffusion, registration and participation, with a within-day ride-hailing model (Figure \ref{fig:Conc-fw}). This model simulates within-day interactions between driver agents, traveller agents and the platform agent.

In this section, we describe the five sub models constituting our approach. We also provide more information about the implementation of the model.

\subsection{Information diffusion}
\cite{rogers1995lessons} argues that the diffusion of information about an innovation is a social process. Individuals seek information from peers to guide the adoption decision, especially from those that have previously adopted the innovation. Information spreading via word-of-mouth is considered to be, to some extent, similar to virus transmission in a network. Hence, many information diffusion models are based on compartment models from epidemics \citep{zhang2016dynamics}. In these models, the population is divided into different classes depending on their current stage of the disease, typically distinguishing susceptible, infectious and recovered agents, although many other compartments are possible \citep{pastor2015epidemic}. One of the main benefits of representing information diffusion with epidemic compartment models is that they do not require the specification of the underlying (social) network, which is typically hard to observe in word-of-mouth communication.  

We assume that an SI model with susceptible (i.e. uninformed) and infectious (i.e. informed) agents suffices. Consider a pool of $N$ potentially interested drivers, of which $I(t-1)$ are informed (i.e. interested or registered) at the start of day $t-1$. If we assume that all uninformed drivers are equally likely to be informed on a given day, then we can formulate the probability for an uninformed driver to be informed at the start of the next day as:  
\begin{equation}
    p^{\mathrm{inform}}(t) = \frac{\beta_{\mathrm{inf}} \cdot I(t-1)}{N} \label{eq:inf-diff} 
\end{equation}
in which $\beta_{\mathrm{inf}}$ represents the average information transmission rate, or more specifically, the probability that information is transmitted in a contact between an informed an uninformed agent multiplied by the average daily number of contacts of agents.

\subsection{Platform registration}
Before informed driver agents can participate, they need to trade off registration costs and participation benefits. In contrast to the approach of \citet{cachon2017role} - one of the few works to represent the registration process in ridesourcing supply - we assume that informed drivers base their registration decision on the average expected income of already registered drivers, rather than on a probability distribution of incomes presented to drivers in advance. Since registration represents a relatively long-term labour decision, we model the decision to be occasional rather than daily. More specifically, we assume that on any given day drivers have a probability $\gamma$ of making a registration decision. 

Drivers register with the platform when the expected earnings from participating exceed the total costs related to participation and registration. Participation cost includes the opportunity cost of the time spent working as well as a potential disutility associated with the driving activity. From hereon, participation cost will be referred to as the reservation wage, a term used in labour economics to define the minimum income level for which drivers are willing to accept specific work \citep{franz1980reservation}. Registration costs, on the other hand, correspond to capital expenses which are independent of participation, such as investment in a vehicle and insurance. We formalize drivers' registration choice with a binary random utility model, in which the registration utility of an informed driver agent $d$ is determined by the net income that drivers expect to collect with participation on the platform, which is defined as the average expected income of already registered drivers $\overline{I_{t}^{\mathrm{exp}}}$ minus a constant penalty $C_d$ to represent capital registration costs. The alternative utility - to remain unregistered - is determined by the reservation wage to represent the time cost of participation on the platform. We apply a logit model with parameter $\beta_{\mathrm{reg}}$, and an error term $\varepsilon_{\mathrm{reg}}$ to account for unknown dynamics in registration choice. The particular utilities and the resulting probability of registration for a driver $d$ on day $t$ are, respectively, formulated as:
\begin{equation}
    U_{dt}^{\mathrm{regist}} = \beta_{\mathrm{reg}} \cdot (\overline{I_{t}^{\mathrm{exp}}} - C_d) + \varepsilon_{\mathrm{reg}}
\end{equation}
\begin{equation}
    U_{dt}^{\mathrm{unregist}} = \beta_{\mathrm{reg}} \cdot W_d + \varepsilon_{\mathrm{reg}}
\end{equation}
\begin{equation}
    p^{\mathrm{regist}}(d,t) = \frac{\gamma \cdot \exp(U_{dt}^{\mathrm{regist}})}{\exp(U_{dt}^{\mathrm{regist}})+\exp(U_{dt}^{\mathrm{unregist}})}
\end{equation}

\subsection{Labour participation}
In the following, we introduce the specification of registered drivers' participation choice, including how drivers anticipate future income based on personal experience.

\subsubsection{Participation choice}
Similar to other studies representing ridesourcing supply \citep{banerjee2015pricing,djavadian2017agent,taylor2018demand,bai2019coordinating}, we model participation based on drivers' expected income and reservation wage. We assume a positive relation between income and labour supply, thus following the neoclassical theory of labour supply \citep{chen2016dynamic,angrist2017uber,xu2020empirical}. Notwithstanding, there are likely to be other factors in play driving participation choice, such as planned activities for the particular day, which are typically difficult to observe. Therefore, in the determination of the utility to participation or to remain idle, next to the reservation wage $W_d$ and expected income $I_{dt}^{\mathrm{exp}}$, we include an error term $\varepsilon_{\mathrm{ptp}}$. We apply a logit model with parameter $\beta_{\mathrm{ptp}}$ and error term $\varepsilon_{\mathrm{ptp}}$ to represent the degree of randomness in the participation choice model, which indicates the significance of non-observed factors influencing the participation choice. The utility and corresponding probability of participating for a driver $d$ on day $t$ are specified as follows: 
\begin{equation}
    U_{dt}^{\mathrm{participate}} = \beta_{\mathrm{ptp}} \cdot I_{dt}^{\mathrm{exp}} + \varepsilon_{\mathrm{ptp}}
\end{equation}
\begin{equation}
    U_{dt}^{\mathrm{idle}} = \beta_{\mathrm{ptp}} \cdot W_d + \varepsilon_{\mathrm{ptp}}
\end{equation}
\begin{equation}
    p^{\mathrm{participate}}(d,t) = \frac{\exp(U_{dt}^{\mathrm{participate}})}{\exp(U_{dt}^{\mathrm{participate}})+\exp(U_{dt}^{\mathrm{idle}})} 
\end{equation}

\subsubsection{Ride-hailing operations}
The financial reward for participation is modeled with the within-day simulation model of the MaaSSim simulator \citep{Kucharski2020MaaSSim}. It allows to capture complex spatiotemporal within-day interactions in ridesourcing between three types of agents: travellers, (participating) drivers and the platform. The following assumptions are made about the operational strategies of these agents in the operational model.

\textit{Driver} agents' labour supply decisions are limited to the extensive margin, i.e. they will work during all hours considered by the within-day model. Drivers will accept all ride requests assigned to them in this time frame. Unassigned drivers do not reposition, instead they remain idle at their drop-off location until assigned to a new request. Driver agents are faced with per-kilometre operating costs $\delta$. 

Each day, a \textit{traveller} agent makes an identical trip for which it requests a ride on the platform. If the time to receive an offer exceeds a threshold $\theta$, the traveller will revoke its ride request. If an offer is received within the tolerance threshold, it will be accepted. Ride offers cannot be cancelled at a later stage.

The \textit{platform} agent offers private rides on a road network with static travel times. It assigns requests to drivers whenever two constraints are met: (1) there are unassigned requests on the platform, and (2) there are idle drivers. It allocates the request-driver pair with the least amount of travel time from the driver's location to the request location. For each transaction, the ridesourcing platform charges a commission rate $\pi$. Ride fares on the platform are comprised of a base fare $f_{\mathrm{base}}$ and per-kilometre fare $f_{\mathrm{km}}$. 

We now specify $Q_{\mathrm{req}}$ as the (virtual) queue of unassigned requests on the platform and $Q_{\mathrm{driver}}$ as the (virtual) queue of idle drivers. $tt_{iu}$ corresponds to the travel time from the location of an idle driver $i\in Q_{\mathrm{driver}}$ to the pick-up location of an unassigned request $u\in Q_{\mathrm{req}}$. The matching function to find the request-driver pair $(u^*,i^*)$ with the least intermediate travel time is then formulated as follows:
\begin{equation}
    (u^*,i^*) = \argmin_{u \in Q_{\mathrm{req}}, i\in Q_{\mathrm{driver}}} tt_{iu}
\end{equation}
The earnings of ridesourcing drivers follow directly from ride fares paid by travellers. If the daily pool of travel requests is denoted as $R$, and the direct distance from request location to destination is denoted as $s_r$, the payout $PO_r$ to a driver for serving a single request $r\in R$ is defined as:
\begin{equation}
    PO_{r} = (f_{\mathrm{base}} + f_{\mathrm{km}} \cdot s_r) \cdot (1 - \pi)
\end{equation}
The total payout $PO_{dt}$ to a driver on a specific day is the sum of the payouts $PO_r$ from requests served by this specific driver on a particular day $t$. Defining $a_{rdt}$ as a binary assignment variable indicating whether driver $d$ picks up request $r$ on day $t$, we can formulate driver's daily payout as:
\begin{equation}
    PO_{dt} = \sum_{r\in R} PO_{r} \cdot a_{rdt}
\end{equation}
The net experienced income of a participating driver $I_{dt}^{\mathrm{act}}$ can now be formulated as:
\begin{equation}
    I_{dt}^{\mathrm{act}} = PO_{dt} - OC_{dt}
\end{equation}
where, in consideration of deadheading distance $DH_{dt}$, $OC_{dt}$ represents a driver's operational costs on day $t$:
\begin{equation}
    OC_{dt} = (\sum_{r\in R} s_r \cdot a_{rdt} + DH_{dt}) \cdot \delta 
\end{equation}

\subsubsection{Learning}
As stated before, participation choice depends on the earnings that are expected on a particular day. Given that the typical ridesourcing driver has limited connections to other drivers \citep{robinson2017making}, anticipated earnings are predominantly based on individual experiences. Considering memory decay \citep{ebbinghaus2013memory} and dynamics in ridesourcing system variables, we cannot assume that drivers weigh all experiences equally. In the absence of empirical evidence for the specification of the learning function in ridesourcing labour supply, we rely on findings from learning in another travel-related context. \cite{bogers2007modeling} demonstrate that conditional on sufficient experience, learning in route choice can be described using a Markov formulation. In this study, we propose a two-phase learning model for driver's perceived income to differentiate learning behaviour by experienced and inexperienced drivers. When the number of days of participation experience $E_{dt}$ exceeds a threshold $\omega$, learning is described with a Markov formulation similar to \cite{bogers2007modeling}. However, when $E_{dt}$ is below $\omega$, drivers compute the unweighted average past income as a proxy for their expected income, to prevent oversensitive and abrupt reactions to the first few experiences. With the actual experienced income on the previous day specified as $I_{d,t-1}^{\mathrm{act}}$, we define the expected income $I_{dt}^{\mathrm{exp}}$ of driver $d$ for day $t$ as:
\begin{equation}
    I_{dt}^{\mathrm{exp}} = (1-\kappa)\cdot I_{d,t-1}^{\mathrm{exp}} + \kappa\cdot I_{d,t-1}^{\mathrm{act}} \label{eq:learning-1}
\end{equation}
in which $\kappa$ represents the weight attributed to the last experience as opposed to all previous experiences, which is formulated as:
\begin{equation}
\kappa = \begin{cases} 
      0 & \hspace{7mm} w_{d,t-1} = 0 \\
      1 / (E_{dt}+1) & \hspace{7mm} 0 < E_{dt} \cdot w_{d,t-1} < \omega \\
      1 / \omega & \hspace{7mm} \text{otherwise} \label{eq:learning-2}
  \end{cases}
\end{equation}
in which $w_{di}$ is a binary variable to indicate whether a driver participated on a past day $i\in\{1,\dots,t-1\}$ and $E_{dt}$ defines the number of days during which the driver has so far gained a participation experience:
\begin{equation}
    E_{dt} = \sum_{i\in\{1,\dots,t-1\}} w_{di} \label{eq:learning-3}
\end{equation}

\subsection{Implementation}
In this subsection, we describe the definitions for convergence and the number of replications in the experiment.

\subsubsection{Simulation framework}
We implement our day-to-day driver model in MaaSSim, an open-source agent-based discrete event simulator of mobility-on-demand operations, programmed in Python \citep{Kucharski2020MaaSSim}. Both supply and demand are represented microscopically. For supply this pertains to the explicit representation of single vehicles and their movements in time and space, while for demand this pertains to exact trip request time and destinations defined at the graph node level. Travel times in the network are precomputed and stored in a skim matrix. 

\subsubsection{Convergence}
A key property of ridesourcing systems is that the size of the fleet may fluctuate on a day-to-day basis. Due to a random component in participation choice, these variations even occur when the system is otherwise in a steady state. To determine whether a ridesourcing system has achieved a steady state, we therefore need to examine other indicators. We argue that a combined analysis of two indicators suffices to establish the convergence of the system. First, there should be few new entrants in the market, i.e. the number of agents with the ability to participate is relatively stable. Second, the degree of learning among registered drivers needs to be minimal, i.e. their expected reward of participation is relatively stable. Together, those criteria imply that the expected fleet size shows limited variations from day to day. The number of drivers that actually decide to participate may still fluctuate due to stochasticity in the participation decision.

We formalise the convergence criteria by checking whether relative day-to-day changes in the number of registered drivers $G_t$ and the expected income of registered drivers $I_{d,t}^{\mathrm{exp}}$ exceed a convergence parameter $\varphi$. The supply evolution process has sufficiently converged when $\varphi$, which is set to approach 0, has not been exceeded on $k$ consecutive days:

\begin{equation}
    \frac{G_{t-j}-G_{t-j-1}}{N_{t-j-1}} \leq \varphi \hspace{5mm},  \forall j\in \{0,1,\dots,k-2,k-1\}
\end{equation}
\begin{equation}
    \frac{|I_{d,t-j}^{\mathrm{exp}}-I_{d,t-j-1}^{\mathrm{exp}}|}{I_{d,t-j-1}^{\mathrm{exp}}} \leq \varphi \hspace{5mm} \forall d\in G_t,  \forall j\in \{0,1,\dots,k-2,k-1\}
\end{equation}
\subsubsection{Replications}
Due to stochastic components in information diffusion, platform registration and participation, we need to replicate the experiment for statistical significance. We determine the number of required iterations $R(m)$ based on a number of initial replications $m$, with a formula commonly used in stochastic traffic simulations \cite{burghout2004note}:
\begin{equation}
    R(m) = \left(\frac{S(m)\cdot t_{m-1,\frac{1-\alpha}{2}}}{\overline{X}(m)\cdot\varepsilon_{\mathrm{repl}}}\right)^2
\end{equation}
where $\overline{X}(m)$ and $S(m)$ are, respectively, the estimated mean and standard deviation of the mean expected income in the population in equilibrium from a sample of $m$ runs, $\varepsilon_{\mathrm{repl}}$ is the allowable percentage error of estimate $\overline{X}(m)$ of the actual mean, and $\alpha$ is the level of significance.

\section{Experimental design}
A series of experiments are constructed for investigating the significance of supply market conditions, platform pricing and service entry barriers in ridesourcing provision. In this section, we introduce the experimental design.

\subsection{Set-up}
We apply the proposed approach to the city of Amsterdam, currently hosting ridesourcing service UberX. It is estimated that in 2019, a total of 8 million taxi or ridesourcing rides took place in Amsterdam, served by 5,000 - 7,000 drivers \citep{gemamsterdam2019}. On an average day, this amounts to approximately 20,000 hailed rides. Considering that it is not likely that all people in Amsterdam potentially interested in driving for a ridesourcing platform actually served at least a single ridesourcing ride in 2019, we assume that the total ridesourcing supply pool in Amsterdam consists of 10,000 drivers. 

Demand is sampled once from a database of rides longer than 2.5 kilometres, generated by the activity-based model Albatross for the Netherlands \citep{arentze2004learning}. It is assumed that travellers are willing to wait five minutes to be matched after requesting their ride, i.e. patience threshold $\theta$ is set to 5. Participating drivers do not make within-day work shift decisions. A single day in the simulation consists of eight hours, corresponding to a typical working day. We simplify the performance of the underlying road network with a universal (constant) traffic speed of 36 km/h on all network links. Ride fares in the experiment are equal to the standard tariffs charged to travellers by Uber in Amsterdam \citep{Uber2020wat}, i.e. a base fare of \euro{1.40} and an additional \euro{1.21} per kilometre. Unlike Uber's pricing model, there is no minimum ride tariff. In the reference scenario that is used throughout the experiment, the commission rate $\pi$ is set to Uber's 25\% \citep{Uber2020track}. As it has been demonstrated that the reservation wage of Uber drivers might be higher or lower than the minimum wage in a given labour market \citep{Chen2019value}, we set the reservation wage $W_d$ in the experiment to \euro{80}, which is close to the minimum daily wage in the Netherlands \citep{GovNL2020minwage}.

We set the information transmission rate $\beta_{\mathrm{inf}}$ to 0.2 so that after around 50 iterations all potential drivers are likely to be informed. Choice model parameters $\beta_{\mathrm{reg}}$ and $\beta_{\mathrm{ptp}}$ are set to 0.2 and 0.1 respectively, representing that unobserved factors are likely to play a larger role in short-term participation, when drivers have more information about the specification of these variables, compared to registration. With $\gamma$ set to 0.2, we expect 20\% of informed drivers to make a registration decision on a given day. The learning threshold $\omega$ is set to 5 days, implying that after five experiences the weight of each new experience in the determination of the expected income has dropped to 0.2, and remains equal afterwards. Convergence parameters $\varphi$ and $k$ are set to 0.01 and 10, respectively.

With each driver assigned a probability of $10 / N$ to be registered at the start of the simulation, we expect an initial registration volume of 10 drivers. Their initial expected income $I^{\mathrm{exp}}_0$ is set to the sum of reservation wage $W_d$ (\euro{80} in the reference scenarios) and the daily share of registration costs $C_d$ (\euro{20} in the reference scenarios). All other driver agents start in the uninformed state.

We empirically establish that the computational load of a single day in the simulation scales directly with the number of requests and vehicles in the system, implying that if we represent the real-world population with a 10\% sample for supply and demand, similar to other studies applying agent-based models in the transportation field \citep{kaddoura2015marginal,bischoff2016autonomous}, we can reduce the total computational load of our experiment by 90\%. Given that we perform a scenario analysis in which each scenario requires multiple replications of our day-to-day simulation approach, we can benefit greatly from the efficiency gain offered by sampling. However, we need evidence that sampling has a limited effect on our simulation results, especially given that ridesourcing may benefit from economies of scale \citep{zha2016economic}. Therefore, we compare the resulting system performance indicators for a 10\% sample of demand and supply to the indicators when we do not apply sampling. Based on three replications for each scenario, we observe that a less efficient matching algorithm in the scenario with sampled supply and demand may lead to a slightly higher average waiting time for travellers, indicating that simulation based on a 10\% sample might lead to slightly overestimated travel times. Remarkably, other performance indicators of the service do not seem to be affected by sampling. The expected income in equilibrium, for example, differs by less than 1\%. Only in the early driver adoption stage, with limited supply, we note a discrepancy in the average income of drivers, which is quickly overcome once supply increases. Our analysis demonstrates that registration and participation volumes scale directly from a 10\% sample to supply and demand levels representing the full population, which indicates that in this case a 10\% sample of supply ($N=1000$) and demand ($M=2000$) is sufficiently large to represent ridesourcing dynamics for the whole city.

When deciding how many replications of the experiment are needed, we allow a relative error $\varepsilon_{\mathrm{repl}}$ of 0.01, based on statistical significance $\alpha$ of 0.01.

\subsection{Scenario design}

\subsubsection{Supply market}
In this part of the experiment, we investigate the extent to which the volume of the pool of potential drivers $N$ is a decisive factor for ridesourcing supply in equilibrium. Compared to the reference scenario (\textit{DP1000} in Table \ref{tab:scenarios}), which assumes a relatively large pool of potential drivers compared to current supply in the network, in alternative scenarios (\textit{DP200} - \textit{DP800}) we test values for $N$ that are smaller, i.e. between 200 and 800 drivers with intervals of 200.

\begin{table}[!ht]
	\tbl{Scenario design} 
		{\begin{tabular}{l c c c c c} \toprule
             Label & $N$ (-) & $W_d$ (\euro) & $\beta_{\mathrm{ptp}}$ & $\pi$ (-) & $C_d$ (\euro) \\ \midrule
             DP200 & 200 & 80 & 0.2 & 0.25 & 20 \\
             DP400 & 400 & 80 & 0.2 & 0.25 & 20 \\
             DP600 & 600 & 80 & 0.2 & 0.25 & 20 \\
             DP800 & 800 & 80 & 0.2 & 0.25 & 20  \\
             DP1000* & 1,000 & 80 & 0.2 & 0.25 & 20 \\ \midrule
             RW50 & 1,000 & 50 & 0.2 & 0.25 & 20  \\
             RW60 & 1,000 & 60 & 0.2 & 0.25 & 20 \\
             RW70 & 1,000 & 70 & 0.2 & 0.25 & 20 \\
             RW80* & 1,000 & 80 & 0.2 & 0.25 & 20 \\
             RW90 & 1,000 & 90 & 0.2 & 0.25 & 20  \\
             RW100 & 1,000 & 100 & 0.2 & 0.25 & 20 \\
             RW110 & 1,000 & 110 & 0.2 & 0.25 & 20 \\ \midrule
             HR0* & 1,000 & $\mathcal{N}$(80,0) & 0.2 & 0.25 & 20 \\
             HR10 & 1,000 & $\mathcal{N}$(80,10) & 0.2 & 0.25 & 20 \\
             HR20 & 1,000 & $\mathcal{N}$(80,20) & 0.2 & 0.25 & 20 \\
             HR30 & 1,000 & $\mathcal{N}$(80,30) & 0.2 & 0.25 & 20 \\ \midrule
             IV005 & 1,000 & 80 & 0.05 & 0.25 & 20 \\
             IV010 & 1,000 & 80 & 0.1 & 0.25 & 20 \\
             IV020* & 1,000 & 80 & 0.2 & 0.25 & 20 \\
             IV050 & 1,000 & 80 & 0.5 & 0.25 & 20 \\
             IV100 & 1,000 & 80 & 1.0 & 0.25 & 20 \\ \midrule
             CF5 & 1,000 & 80 & 0.2 & 0.05 & 20  \\
             CF15 & 1,000 & 80 & 0.2 & 0.15 & 20 \\
             CF25* & 1,000 & 80 & 0.2 & 0.25 & 20 \\
             CF35 & 1,000 & 80 & 0.2 & 0.35 & 20 \\
             CF45 & 1,000 & 80 & 0.2 & 0.45 & 20 \\
             CF55 & 1,000 & 80 & 0.2 & 0.55 & 20 \\ \midrule
             RC0 & 1,000 & 80 & 0.2 & 0.25 & 0 \\
             RC10 & 1,000 & 80 & 0.2 & 0.25 & 10 \\
             RC20* & 1,000 & 80 & 0.2 & 0.25 & 20 \\
             RC30 & 1,000 & 80 & 0.2 & 0.25 & 30 \\
             RC40 & 1,000 & 80 & 0.2 & 0.25 & 40 \\\bottomrule
		\end{tabular}} 
		\tabnote{*Reference scenario}
		\label{tab:scenarios}
\end{table}

Another supply market condition that is expected to affect emergent ridesourcing supply is the reservation wage, which may be high or low depending for example on the ease of access to alternative sources of income \citep{baron2018disruptive,Chen2019value}. First, we examine six alternative scenarios in which the reservation wage is considered to be homogeneous across the population of drivers. With these scenarios, labeled \textit{RW50} - \textit{RW110} in Table \ref{tab:scenarios}, we cover the range of reservation wages from \euro 50 to \euro 110. Then, we consider three additional scenarios with heterogeneity in reservation wage $W_d$, to represent that the opportunity cost of ridesourcing participation may vary across the population due to uneven opportunities in the labour market. We represent the heterogeneity in $W_d$ with a normal distribution in which the mean is equal to the homogeneous reservation wage value from the reference scenario (\euro80). In scenarios \textit{HR0} - \textit{HR30}, we test the effect of reservation wage heterogeneity on ridesourcing supply with four values for the standard deviation of the reservation wage distribution: \euro0 (i.e. homogeneous reservation wage), \euro10, \euro20 and \euro30. 

Since participation in our approach is modelled with a probabilistic participation choice model, we can also investigate how opportunistic behaviour in labour supply affects ridesourcing supply levels. We do this by varying the participation logit model parameter $\beta_{\mathrm{ptp}}$, representing the relative weight that drivers assign to income as opposed to other, in our model unobserved, variables. Lacking empirical evidence for the value of $\beta_{\mathrm{ptp}}$, in scenarios \textit{IV005} - \textit{IV100} we test a relatively large range of values: from 0.05 to 1.0.

\subsubsection{Platform pricing}
The main instrument that ridesourcing platforms hold to steer supply is their pricing strategy, including the ride fare structure and the commission rate, i.e. the proportion of each transaction retained by the platform. We investigate the implications of price settings in the ridesourcing market for drivers and travellers, accounting for the dynamics related to supply, by analysing a series of scenarios covering a relatively large range of commission rates $\pi$: from a limited 5\% to more than half of the ride fare, 55\%, with intervals of 10\%. The scenarios are included in Table \ref{tab:scenarios} as scenarios \textit{CF5} - \textit{CF55}.

\subsubsection{Entry barriers}
Ridesourcing uptake - and potentially excessive competition - on the supply side may partially be accredited to low entry barriers \citep{rayle2016just}. On the other hand, a lack of capital participation costs may also lead to less frequent participation \citep{hall2018analysis}. Hence, we investigate the effect of financial entry barriers, such as a taxi license, on emergent ridesourcing supply. We examine five scenarios for which we vary the registration cost parameter $C_d$, which represents costs that are sunk in participation but not in registration. We consider two extreme scenarios, one in which capital costs are absent, and one in which capital costs add up to half the reservation wage (\euro40). In the three intermediate scenarios, the relative penalty for registration amounts either \euro10, \euro20 or \euro30. In Table \ref{tab:scenarios}, the scenarios are labeled \textit{RC0} - \textit{RC40}.

\subsection{User equilibrium optimality}
Unlike transportation services in which drivers are employed by the service provider, supply in ridesourcing is a decentralised process centered around the labour decisions of individual drivers. So far, we have considered how to test the effect of labour market characteristics, platform policies and entry barriers on ridesourcing supply, but not yet how the emerging user equilibria compare to supply if controlled by a central service provider or organisations representing the interests of travellers and drivers. Specifically, we investigate the optimality of decentralised ridesourcing supply from three different perspectives:

\begin{itemize}
    \item \textit{Service provider (platform)}: Aims to maximise the profit from collecting a fee from each transaction between travellers and drivers
    \item \textit{Traveller union}: Representing the interests of travellers, it aims to minimise travel times and rejected requests. We formalize this objective with a value of time of \euro{8/h}, which was found to be the average value for travellers in the Netherlands \citep{RWS2020kengetallen}, and assigning a penalty of \euro8 for each rejected request.
    \item \textit{Driver union}: Representing the interests of the driver community, it aims to maximise total driver surplus in the system. The surplus for an individual driver is defined as the difference between its actual earnings ${I_{dt}^{\mathrm{act}}}$ and reservation wage $W_d$ \citep{Chen2019value}.
\end{itemize}

We search for the optimal fleet size for the three different parties by performing a brute-force search, testing their respective objective functions for a single day assuming various participation volumes. We test values around the user equilibrium in the base scenario: from 20 to 300 participating drivers in steps of 20, i.e. $m = [20,40,\dots,280,300]$.

\section{Results}
We analyse the results of our experiments focusing on the evolution process of ridesourcing and specifically the role of the supply market and pricing policy. Table \ref{tab:results} contains the comprehensive list of KPIs on day 200 of our iterative simulation, when all replication runs have converged to an equilibrium.

\begin{table}[ht!]
    \sisetup{round-mode=places}
    \tbl{KPIs in equilibrium for all scenarios}
        {\begin{tabular}{l c c c c c c c c c c c} \toprule
            Label & \rot{Informed drivers} & \rot{Registered drivers} & \rot{Participating drivers} & \rot{\shortstack[l]{Expected income, \\ mean (\euro)}} & \rot{\shortstack[l]{Expected income, \\ std. among drivers (\euro)}} & \rot{\shortstack[l]{Experienced income, \\ mean (\euro)}} & \rot{\shortstack[l]{Experienced income, \\ std. among drivers (\euro)}} & \rot{Satisfied requests (\%)} & \rot{Average waiting time (s)} & \rot{Daily platform profit (\euro)} & \rot{\shortstack[l]{Convergence criterion \\ satisfied (day)}} \\ \midrule
            DP200  &     200 &     198 &       125 &          85.96 &         10.49 &         89.36 &        19.93 &         100.0 &          162.8 &        5217 & 47.0 \\
            DP400  &     400 &     301 &       136 &          77.56 &         10.62 &         82.25 &        20.75 &         100.0 &          153.6 &        5217  & 58.8 \\
            DP600  &     600 &     350 &       140 &          75.04 &         10.57 &         80.38 &        21.15 &         100.0 &          151.1 &        5217 & 68.2 \\
            DP800  &     800 &     377 &       142 &          73.79 &         10.77 &         79.48 &        20.85 &         100.0 &          150.5 &        5217 & 61.3 \\
            DP1000 &    1000 &     425 &       145 &          71.83 &         10.29 &         77.58 &        21.47 &         100.0 &          148.8 &        5217 & 64.8  \\\midrule
            RW50  &    1000 &     572 &       228 &          45.11 &         11.38 &         51.19 &        21.27 &         100.0 &          127.0 &        5351 & 58.1 \\
            RW60  &    1000 &     504 &       192 &          54.17 &         11.33 &         60.63 &        22.49 &         100.0 &          137.7 &        5351 & 68.5 \\
            RW70  &    1000 &     454 &       167 &          63.34 &         11.21 &         69.41 &        22.02 &         100.0 &          145.5 &        5351 & 62.0 \\
            RW80  &    1000 &     422 &       147 &          72.42 &         11.03 &         78.36 &        21.51 &         100.0 &          153.5 &        5351 & 64.2 \\
            RW90  &    1000 &     396 &       133 &          81.53 &         10.81 &         86.49 &        21.92 &         100.0 &          164.2 &        5351 & 74.2 \\
            RW100 &    1000 &     368 &       118 &          90.98 &         10.21 &         96.56 &        20.89 &         100.0 &          185.3 &        5351 & 75.8 \\
            RW110 &    1000 &     339 &       106 &         100.33 &          9.76 &        105.14 &        17.58 &          99.7 &          227.9 &        5334 & 71.0 \\\midrule
            HR0  &    1000 &     398 &       143 &          71.56 &         13.62 &         78.98 &        23.22 &         100.0 &          152.8 &        5230 & 70.0 \\
            HR10 &    1000 &     382 &       149 &          67.46 &         14.23 &         75.76 &        24.07 &         100.0 &          149.2 &        5230 & 55.4 \\
            HR20 &    1000 &     393 &       168 &          60.95 &         14.01 &         67.51 &        24.12 &         100.0 &          138.0 &        5230 & 52.4 \\
            HR30 &    1000 &     410 &       188 &          55.43 &         14.49 &         60.56 &        22.90 &         100.0 &          130.5 &        5230 & 53.6 \\\midrule
            IV005 &    1000 &     401 &       158 &          70.36 &         10.67 &         72.51 &        22.33 &         100.0 &          150.0 &        5312 & 56.5 \\
            IV010 &    1000 &     415 &       148 &          72.86 &         10.32 &         77.48 &        20.77 &         100.0 &          155.8 &        5312 & 60.3 \\
            IV020 &    1000 &     432 &       137 &          74.15 &         10.30 &         83.11 &        20.94 &         100.0 &          161.3 &        5312 & 65.5 \\
            IV050 &    1000 &     482 &       125 &          75.86 &         10.63 &         90.79 &        21.00 &         100.0 &          173.8 &        5312 & 68.5 \\
            IV100 &    1000 &     538 &       120 &          77.67 &         10.51 &         93.86 &        20.70 &         100.0 &          176.9 &        5312 & 91.7 \\\midrule
            CF5 &    1000 &     472 &       181 &          73.23 &         16.91 &         85.61 &        30.94 &         100.0 &          129.2 &        1042 & 56.4 \\
            CF15 &    1000 &     441 &       163 &          72.95 &         13.11 &        82.15 &        25.47 &         100.0 &          137.5 &        3125 & 62.4 \\
            CF25 &    1000 &     400 &       144 &          72.85 &         10.24 &         77.94 &        20.29 &         100.0 &          148.0 &        5195 & 72.6 \\
            CF35 &    1000 &     356 &       120 &          72.25 &          7.61 &         75.25 &        14.24 &         100.0 &          166.1 &        7265 & 73.4 \\
            CF45 &    1000 &     240 &        70 &          70.70 &          3.88 &         72.04 &         8.26 &          86.7 &          363.3 &        7962 & 77.0 \\
            CF55 &    1000 &      52 &        11 &          67.03 &          2.08 &         67.29 &         4.14 &          16.9 &          194.1 &        1751 & 18.4 \\\midrule
            RC0  &    1000 &     885 &       168 &          62.52 &         10.73 &         69.30 &        22.75 &         100.0 &          143.4 &        5394 & 68.3 \\
            RC10 &    1000 &     632 &       157 &          66.57 &         10.95 &         74.35 &        21.72 &         100.0 &          148.6 &        5388 & 63.0 \\
            RC20 &    1000 &     429 &       149 &          72.12 &         11.19 &         78.13 &        20.49 &         100.0 &          152.9 &        5384 & 68.7 \\
            RC30 &    1000 &     287 &       138 &          78.87 &         11.08 &         84.04 &        20.53 &         100.0 &          158.0 &        5384 & 60.0 \\
            RC40 &    1000 &     205 &       128 &          85.84 &         11.24 &         89.75 &        19.09 &         100.0 &          164.1 &        5383 & 55.7 \\\bottomrule
        \end{tabular}}
    \label{tab:results}
\end{table}

\subsection{Phases in ridesourcing provision}\label{subsect:phases}
\begin{figure}[]
  \centering
  \includegraphics[width=\textwidth]{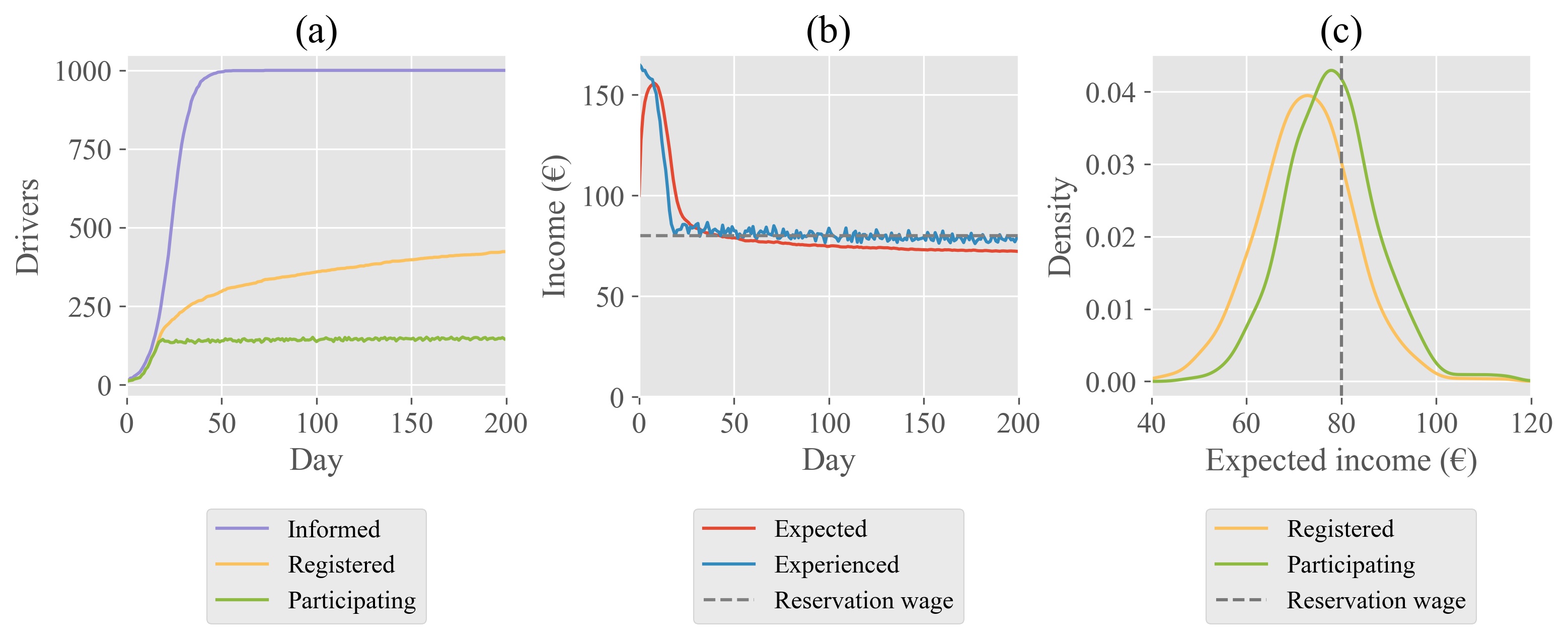}
  \caption{(a) The evolution of the number of registered, informed and participating drivers, (b) the evolution of the average expected and experienced income, and (c) the distribution of expected income for participating drivers versus registered drivers in equilibrium} \label{fig:base-scn}
\end{figure}

In this subsection, we examine the evolution of ridesourcing supply and the implications for suppliers specifically for one of the reference scenarios, $\textit{RW80}$ (Figure \ref{fig:base-scn}a,b). In accordance with the specification of the information diffusion process, all 1,000 driver agents are eventually informed about the existence of the service. In equilibrium, considering multiple simulation iterations, after 200 days on average less than half of those agents (420) are registered, of which on a typical day approximately a third participate (145 drivers). We identify five phases in the evolution process:

\begin{enumerate}
    \item \textit{Day 0 - 10}: Due to a lack of information, few driver agents have registered, meaning participation is low as well. Participating drivers profit from a lack of competition and can make a high profit.
    \item \textit{Day 10 - 20}: Information transmission speeds up. Informed drivers are likely to register as they observe a high average income. Participation increases rapidly, leading to a collapse in the experienced income. Drivers start to learn that their anticipated income may not be feasible.
    \item \textit{Day 20 - 50}: Information diffusion continues. Drivers further downscale their income expectation based on new participation experiences. As a result of the drop in expected income, the average driver participates less frequently. The number of registered drivers still increases, albeit at a slower pace than before. As a consequence, the total participation volume increases marginally, leading to a further decrease in the experienced income level.
    \item \textit{Day 50 - 100}: All drivers are now informed. Registration continues at a decreasing pace, yet participation increases only marginally since individual drivers participate less frequently, as a result of the continuing decrease in the average expected income.
    \item \textit{Day 100 - 200}: Equilibrium is reached. Registrations and the decrease in experienced and expected income are now limited. Participation remains constant over time.
\end{enumerate}

There are two aspects in Figure \ref{fig:base-scn}b worth highlighting. First, the average expected income of drivers converges to a value below the average experienced income. Figure \ref{fig:base-scn}c provides an explanation for the discrepancy in expected and experienced income: drivers with a low expected income are relatively unlikely to participate compared to drivers with a higher expected income, and consequently less likely to 'update' their expected income based on a new (likely more positive) driving experience. Convergence is reached when the average experienced income is equal to the average expected income of participating drivers, which is higher than the average expected income of all - also non-participating - drivers. Second, the presented evolution process demonstrates that, when we assume that variables other than expected income play a role in participation choice, the average daily income of participating drivers on the platform may converge to a value below the reservation wage (Figure \ref{fig:base-scn}b). This can be attributed to unobserved variables in participation, like scheduled activities for a given day, which cause a significant group of drivers to work even when their experienced income is below the reservation wage (Figure \ref{fig:base-scn}c). In fact, more than half of the drivers that participate on a given day in the equilibrium expect to earn less than their reservation wage. This finding emphasizes that the main value of a ridesourcing service may be found in the flexibility it offers, as suggested also by \cite{Chen2019value}, rather than in providing a satisfactory level of income over a longer period of time.

\subsection{Supply market conditions}
In this subsection, we present the effect of the size of the driver pool, the reservation wage and unobserved variables in participation on dynamic ridesourcing provision. The information diffusion process is not affected in scenarios, except for those with an alternative size of the driver pool (see Equation \ref{eq:inf-diff}).

\subsubsection{Driver pool}
\begin{figure}[]
  \centering
  \includegraphics[width=\textwidth]{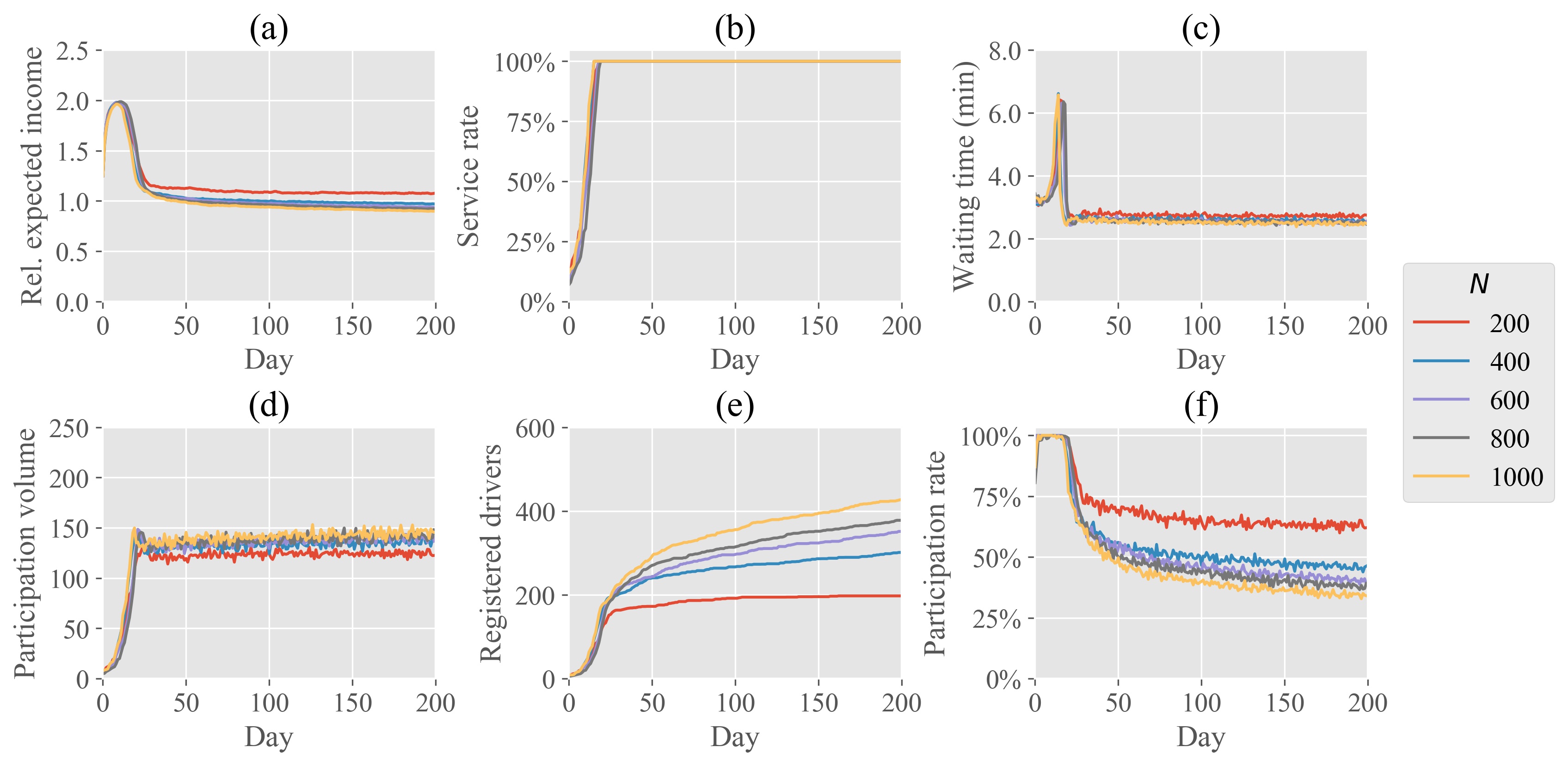}
  \caption{The effect of the size of the driver pool on the evolution of (a) the expected income of registered drivers as ratio of their reservation wage, (b) the share of requests that are satisfied, (c) the average waiting time for pick-up for travellers, (d) daily participation volumes, (e) the total number of registered drivers, and (f) the share of registered drivers that participate}\label{fig:res_pool-six}
\end{figure}

When the pool of drivers is limited to 200 (scenario \textit{DP200} in Table \ref{tab:scenarios}), we find that an equilibrium state is reached around day 50 (Table \ref{tab:results}). In this state, nearly all potential drivers have registered (Figure \ref{fig:res_pool-six}e) and the participation frequency is fairly stable at a high level (Figure \ref{fig:res_pool-six}f). When the pool of potential drivers is larger, there are still unregistered drivers left around this time in the simulation, of whom a part decides to sign up in a later phase. This explains why the transition process takes longer in our experiment when the pool of potential drivers is large.

For 200 potential drivers, we find an equilibrium average expected income for registered drivers that exceeds their reservation wage by nearly 10\%. In all other scenarios, representing supply markets of 400 potential drivers or more, the average drivers fails to match the reservation wage, falling short by 5 - 10\% (Figure \ref{fig:res_pool-six}a). It is striking that there seems to be little difference in service performance when a supply market consists of 1000 drivers as opposed to 400 drivers. In both cases, after approximately 25 iterations supply is sufficient to saturate the market and serve all requests in the system (Figure \ref{fig:res_pool-six}b), without a significant difference in the average waiting time for travellers (Figure \ref{fig:res_pool-six}c). Figure \ref{fig:res_pool-six}d shows that the similarity in travellers' level of service follows directly from the daily participation volume, which is approximately equal in both scenarios. Apparently, 600 additional potential drivers in the supply market only yield around 125 more registrations around the 200th day (Figure \ref{fig:res_pool-six}e), while those that are registered also participate less frequently when the potential supply market is large (Figure \ref{fig:res_pool-six}f), on average 34\% versus 46\% of the days.

The finding that ridesourcing supply converges to an invariant participation volume for different sizes of the labour supply market, as long as the total supply volume is relatively high compared to demand, demonstrates the existence of a balancing effect in ridesourcing supply. In such a market, the frequency of participation compensates for the size of the pool of registered drivers, which means that negative consequences related to oversupply have an inherent upper bound. Notwithstanding, in this upper bound, expected income may be below the reservation wage. Only when the size of the supply market is limited to a value close to the invariant participation volume when the supply market is sufficiently large, we find expected income to exceed the reservation wage. This resonates with the introduction of supply caps, implemented for example in New York City, in raising ridesourcing drivers' average income. The results also show that travellers may not suffer much from a supply cap, at least as long as the cap is set to a sensible level.

\subsubsection{Homogeneous reservation wage}
\begin{figure}[]
  \centering
  \includegraphics[width=\textwidth]{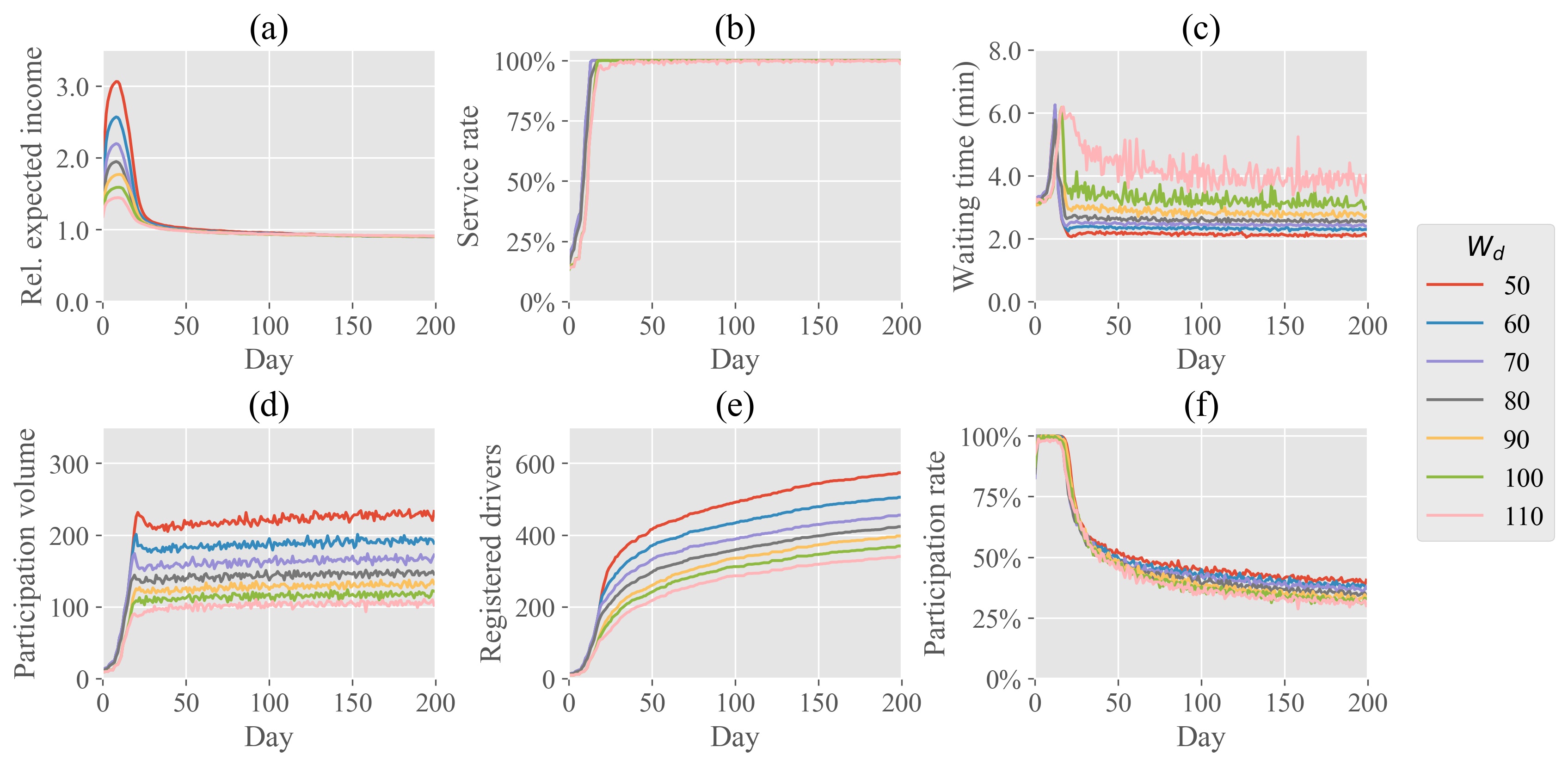}
  \caption{The effect of (homogeneous) reservation wage on the evolution of (a) the expected income of registered drivers as ratio of their reservation wage, (b) the share of requests that are satisfied, (c) the average waiting time for pick-up for travellers, (d) daily participation volumes, (e) the total number of registered drivers, and (f) the share of registered drivers that participate}\label{fig:res_wage-six}
\end{figure}

Based on our experiments, the reservation wage of potential drivers has a minor effect on the duration of the transition process. The equilibrium condition is reached marginally more quickly when the reservation wage of drivers is low (Table \ref{tab:results}). This may be caused by more registrations in an early phase of the evolution (Figure \ref{fig:res_wage-six}e) due to lower labour opportunity costs. While there are still new registrations in a later phase, the relative increase in the size of the pool of registered drivers is low compared to scenarios with higher reservation wages.

Remarkably, we find that in equilibrium the ratio between expected income and reservation wage is constant for various reservation wages (Figure \ref{fig:res_wage-six}a), slightly under 1. It means that as the reservation wage in a market increases, the expected income in equilibrium increases proportionally. The effect of reservation wage on the level of service for travellers seems to be limited. Even in scenario \textit{RW110}, in which labour costs are least favourable for supply, i.e. the reservation wage equals 110 euros, supply is sufficient to serve all requests (Figure \ref{fig:res_wage-six}b), albeit travellers are confronted with longer travel times than in scenarios with a lower cost of labour (Figure \ref{fig:res_wage-six}c). The additional waiting time is, however, limited to a maximum of two minutes and thereby fairly limited. The differences in waiting time stem from participation volumes that vary between 100 and 230 for different specifications of the reservation wage (Figure \ref{fig:res_wage-six}d). Lower participation when labour supply is costly results both from fewer registrations (Figure \ref{fig:res_wage-six}e) and less frequent participation among those registered (Figure \ref{fig:res_wage-six}f).

The results imply that a weak labour market, associated with low reservation wages, leads to reduced income levels for suppliers in the ridesourcing market, because new suppliers are attracted to the market as a result of a lack of alternative employment, creating competition for pick-ups. Ridesourcing providers on the other hand can potentially profit from the inflow of supply in times of economic recession by means of reduced waiting time for travellers, which may attract new demand, or alternatively, by giving them the opportunity to increase the commission rate without sacrificing the level of service for travellers.

\subsubsection{Heterogeneous reservation wage}
It can be expected that the minimum income that drivers want to collect with ridesourcing participation is not equal for all drivers, for example because some drivers have better access to alternative employment than others. To capture reservation wage heterogeneity, one of the set of scenarios included in our experiment is directed at investigating ridesourcing supply for different reservation wage distributions, with the same mean $\mu$ as the reference scenario but different standard deviations $\sigma$. 

\begin{figure}[]
  \centering
  \includegraphics[width=\textwidth]{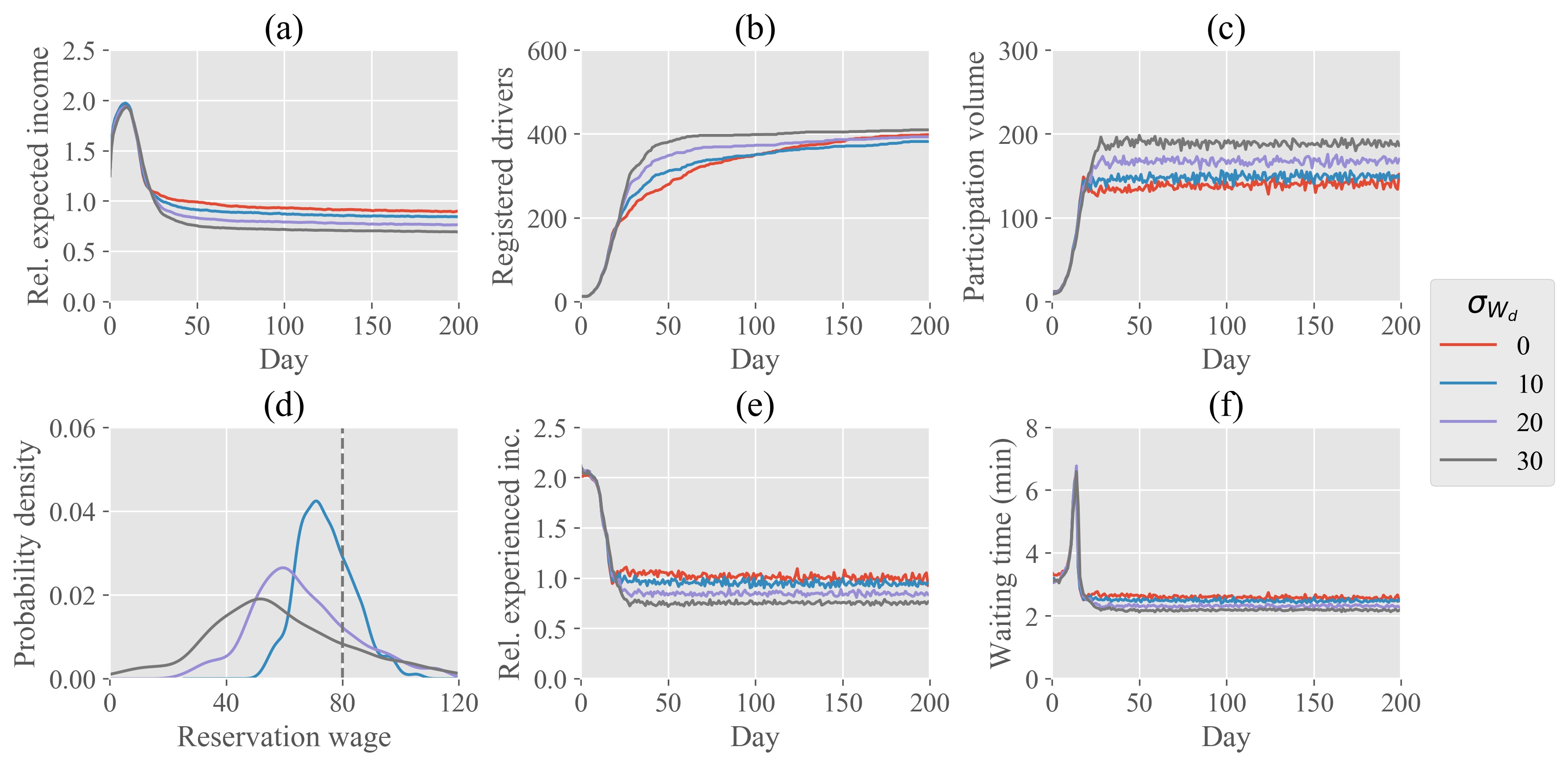}
  \caption{The effect of heterogeneity in reservation wage on (a) the evolution of the average expected income of registered drivers as ratio of their reservation wage, (b) the evolution of the total number of registered drivers, (c) the evolution of daily participation volumes, (d) the probability density function of reservation wage for registered drivers, (e) the evolution of average experienced income of participating drivers as ratio of their reservation wage and (f) the evolution of the average waiting time for pick-up for travellers}\label{fig:het_res_wage-six}
\end{figure}

Figure \ref{fig:het_res_wage-six}a shows that when there is a lot of variation in drivers' reservation wage (scenario \textit{HR30}), the expected income of registered drivers in equilibrium is relatively low. Yet, a high value for $\sigma$ does not seem to lead to a slower registration process (Figure \ref{fig:het_res_wage-six}b). In fact, participation appears to be higher with strong heterogeneity in the reservation wage (Figure \ref{fig:het_res_wage-six}c). Figure \ref{fig:het_res_wage-six}d demonstrates that in such a scenario, a relatively high share of registered drivers has a low reservation wage, meaning that they are relatively like to supply labour on a given day, even when they expect a low income. It explains also why registration (Figure \ref{fig:het_res_wage-six}b) peaks early in a scenario with high $\sigma$: drivers with a reservation wage below the mean benefit significantly from registration and are thus relatively likely to register. Due to the quick influx of drivers and the fact that drivers that are still unregistered have a relatively high reservation wage, registrations then slow down quickly. High participation volumes in scenarios with strong heterogeneity result in a low average income for drivers in the system (Figure \ref{fig:het_res_wage-six}e) and slightly lower waiting times for travellers (Figure \ref{fig:het_res_wage-six}f). 

The results imply that with a high degree of inequality in the labour market, ridesourcing markets may be flooded with drivers with limited labour opportunities elsewhere. Due to their weak position in the labour market, they are willing to work for ridesourcing platforms even when wages are low, providing competition for other participating drivers. Our experiment demonstrates that high participation may only yield limited benefits in terms of the average waiting time for travellers, while the income for drivers may be significantly lower than in scenarios with lower participation. We conclude that especially in labour markets characterised by large inequalities supply caps may be necessary to guarantee a socially desired minimum income for ridesourcing drivers. 

\subsubsection{Unobserved variables in participation}
\begin{figure}[]
  \centering
  \includegraphics[width=\textwidth]{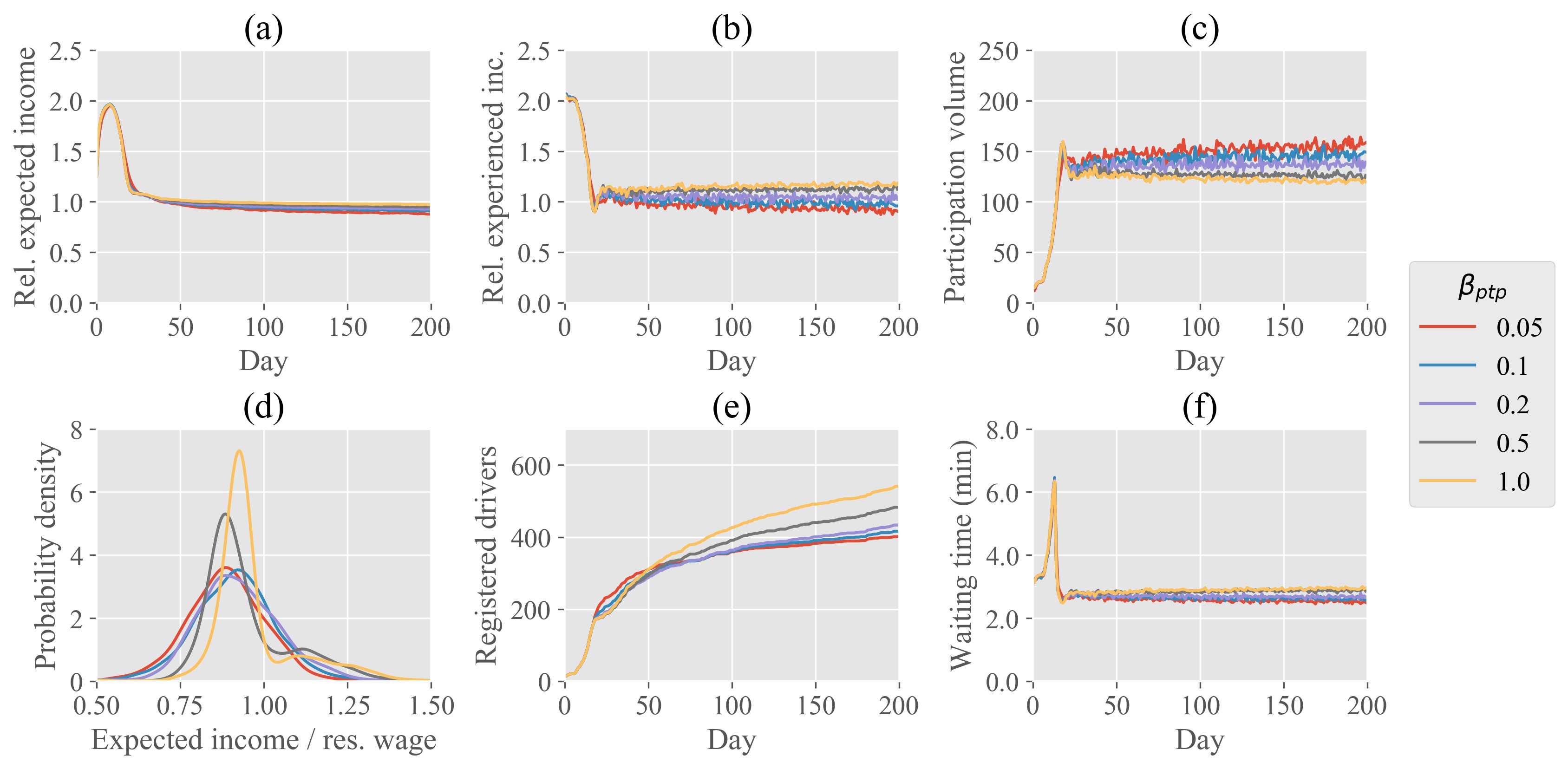}
  \caption{The effect of the valuation of income in participation choice on (a) the evolution of the average expected income of registered drivers as ratio of their reservation wage, (b) the evolution of average experienced income of participating drivers as ratio of their reservation wage, (c) the evolution of daily participation volumes, (d) the probability density function of expected income (as ratio of their reservation wage) for registered drivers, (e) the evolution of the total number of registered drivers and (f) the evolution of the average waiting time for pick-up for travellers}\label{fig:ptcp_beta-six}
\end{figure}

Choice parameter $\beta_{\mathrm{ptp}}$ represents the value drivers attach to income as opposed to other variables in participation decisions. A low $\beta_{\mathrm{ptp}}$ indicates that drivers supply labour to the platform more opportunistically, potentially working one day but not the next even when the income they anticipate is the same. Our results show that while $\beta_{\mathrm{ptp}}$ has a limited effect on the average expected income of drivers registered with a platform (Figure \ref{fig:ptcp_beta-six}a), there is a clear difference in the average actual income generated by participating drivers (Figure \ref{fig:ptcp_beta-six}b). The reason for this discrepancy is that in the scenario with the highest value for $\beta_{\mathrm{ptp}}$ (scenario \textit{IV100}), despite a slightly higher average expected income, on average approximately 40 fewer drivers actually decide to participate compared to the scenario with lowest $\beta_{\mathrm{ptp}}$ (Figure \ref{fig:ptcp_beta-six}c). Figure \ref{fig:ptcp_beta-six}d provides an explanation for this phenomenon. With expected income as the dominant variable for participation when $\beta_{\mathrm{ptp}}$ is high, a driver that expects to make an income just below their reservation wage is relatively unlikely to participate, and consequently, also to update its income expectation based on new, potentially more positive, experiences. In this scenario, drivers confronted with a negative driving experience are therefore less likely to participate thereafter compared to scenarios with a lower value of $\beta_{\mathrm{ptp}}$, resulting in a large group of 'dissatisfied' drivers with an income just below the reservation wage, but ultimately also in a (relatively small) group of drivers profiting from the lack of competition when it comes to serving rides. Participating drivers in this scenario earn on average approximately 15\% more than their reservation wage, compared to 10\% less in the scenario with a $\beta_{\mathrm{ptp}}$ of 0.05. The average waiting time for travellers is, however, also highest in this scenario (Figure \ref{fig:ptcp_beta-six}f).

Due to slightly higher expected earnings when income is the dominant factor in the participation decision, more unregistered drivers decide to sign up in later phases of the transition process (Figure \ref{fig:ptcp_beta-six}e) compared to scenarios in which $\beta_{\mathrm{ptp}}$ is low. Hence, the equilibrium market state is achieved more quickly when drivers attribute more value to variables other than income.

To summarize, if we assume that income is not the sole explanatory variable for participation, in line with what is suggested by early research on labour supply of ridesourcing drivers \cite{Chen2019value}, the average income for participating drivers in a ridesourcing system is likely to turn out relatively low, since every day a portion of drivers is willing to participate for a wage below their reservation wage, increasing competition for supply in the system. This implies that, in such a scenario, the ridesourcing service may be valuable for drivers wishing to supply labour flexibly, utilising the service for example only on days without planned activities or other work, but less so for drivers using the platform as a replacement for a full-time job.

\subsection{Platform policies}
\begin{figure}[] 
  \centering
  \includegraphics[width=\textwidth]{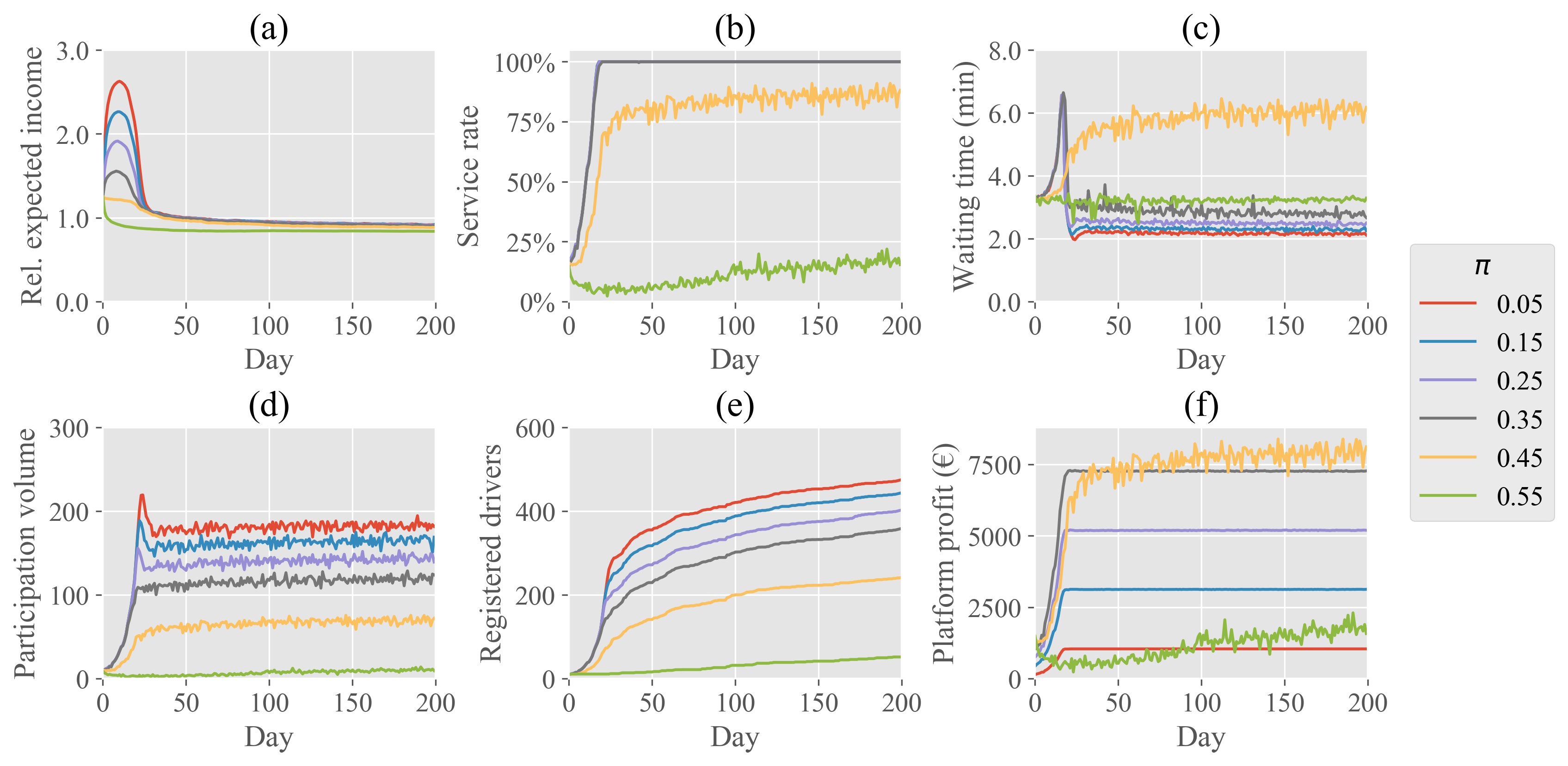}
  \caption{The effect of platform commission rate on the evolution of (a) expected income of registered drivers as ratio of their reservation wage, (b) the share of requests that are satisfied, (c) the average waiting time for pick-up for travellers, (d) daily participation volumes, (e) the total number of registered drivers, and (f) daily platform profit}\label{fig:comm_fee-six}
\end{figure}

We observe that a lower commission allows for higher earnings in early transition phases (Figure \ref{fig:comm_fee-six}a), which convinces more potential drivers to register in this time frame (Figure \ref{fig:comm_fee-six}e). After increased supply-side competition has brought earnings down, the number of new registrations slows down in all scenarios. In scenarios in which initially many drivers register, the relative increase in the size of the pool of registered drivers is lower than in scenarios in which fewer drivers registered. This means that these markets end up in an equilibrium state more quickly. This trend applies however only up to a certain point. When commissions are increased further, corresponding to a commission rate of 55\% in the experiment, hardly any drivers will register at all. In that case, the market equilibrium is achieved very quickly.

Interestingly, we find that the expected income of drivers in equilibrium is hardly affected by the commission fee that is charged by the platform (Figure \ref{fig:comm_fee-six}a). A commission rate of 55\% (scenario \textit{CF55}) yields an expected driver income which is not more than 10\% lower than when the commission rate is set to only 5\%. Ridesourcing users, on the other hand, can strongly be affected by the platform commission rate. The additional inconvenience is fairly limited when the commission rate is set to 35\% as opposed to 5\%, inducing an average additional waiting time of less than one minute. However, with a commission rate of 45\% or 55\%, a part of the requests needs to be rejected and the waiting time of the remaining travellers is significantly longer (Figure \ref{fig:comm_fee-six}b, \ref{fig:comm_fee-six}c). In fact, when the commission rate is 55\%, only 20\% of requests can be satisfied in equilibrium. Figures \ref{fig:comm_fee-six}d and \ref{fig:comm_fee-six}e demonstrate that supply adjusts itself to the commission rate that is in effect, which provides an explanation for why income levels are largely unaffected, while the level of service for an average ride strongly deteriorates. In this particular experiment, a commission rate of 45\% appeared to be the optimal strategy for the ridesourcing provider, generating approximately 8,000 euros per day in equilibrium (Figure \ref{fig:comm_fee-six}f).

These findings demonstrate that, profit-wise, the collection of a higher share per request may outweigh revenue loss from not being able to serve all incoming requests. This implies that profit maximisation in ridesourcing provision may come at the expense of travellers, who are exposed to longer waiting times and a higher probability of being rejected altogether. Interestingly, ridesourcing drivers are hardly affected by strategical platform behaviour relating to the commission rate, since driver registration is slower when commission rates are high. At the same time, we observe that within a certain range, platform profit can be vastly improved without significantly affecting riders in the system. Our experiment shows that a non-optimal pricing strategy in terms of profit (in the experiment a commission rate of 35\% as opposed to 45\%), may result in near-optimal platform profit and driver income, with a much improved level of service for travellers. Thus, it might be worthwhile for authorities to consider regulating the commission rate while considering its consequences for service affordability.

\subsection{Entry barriers}
\begin{figure}[] 
  \centering
  \includegraphics[width=\textwidth]{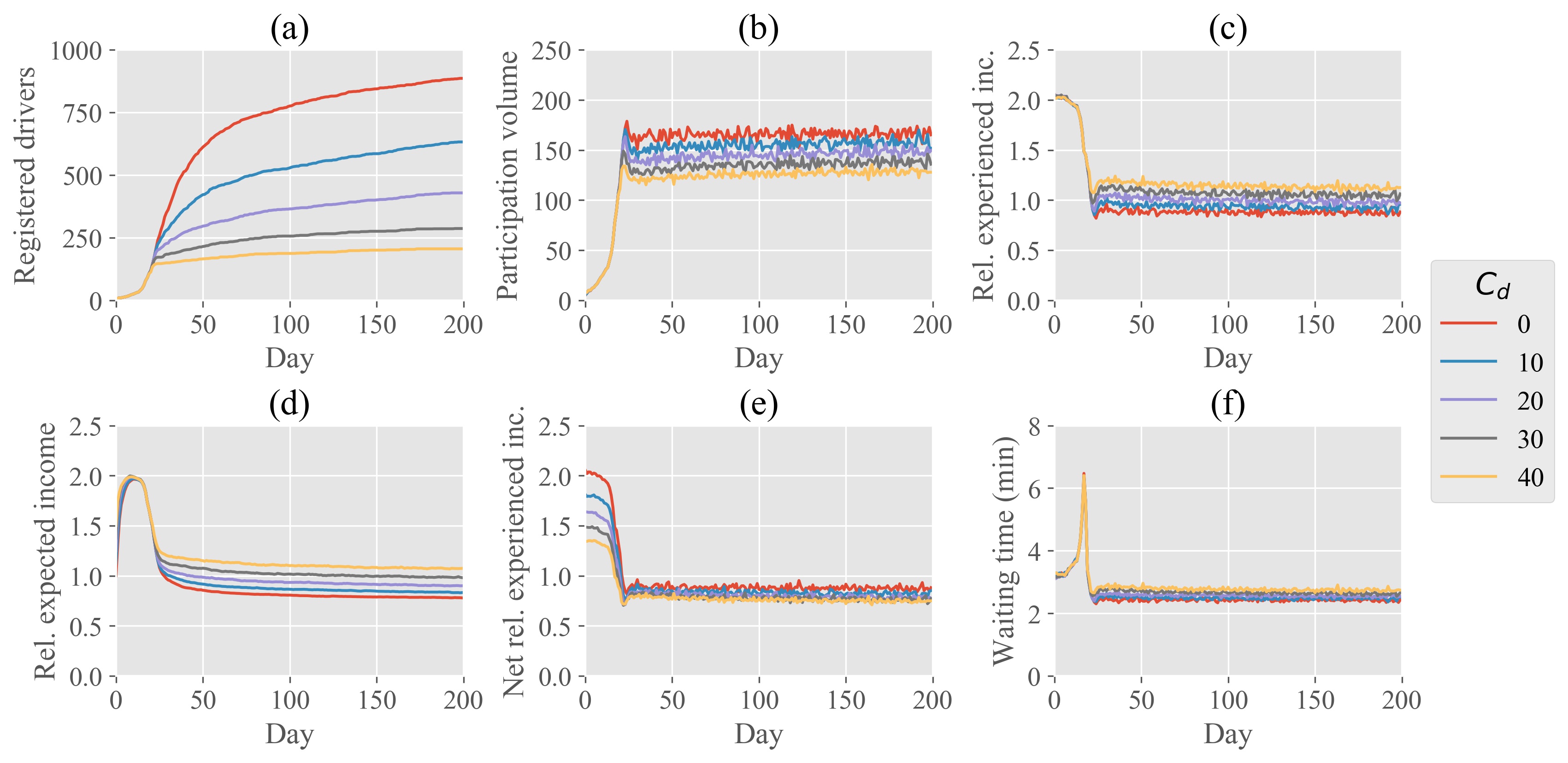}
  \caption{The effect of registration costs on the evolution of (a) the average expected income of registered drivers as ratio of their reservation wage, (b) the total number of registered drivers, (c) daily participation volumes, (d) average experienced income of participating drivers as ratio of their reservation wage, (e) average experienced income of participating drivers as ratio of the sum of their reservation wage and daily share of registration costs and (f) the average waiting time for pick-up for travellers}\label{fig:reg-costs-six}
\end{figure}

The need for vehicle, insurance and medallion acquisition may prevent interested drivers from registering with a ridesourcing platform. In some markets, these factors are more prevalent than in others. We mimic markets with different registration regimes by varying registration cost parameter $C_d$. We find that when registration costs are high, indeed, significantly fewer drivers will register with a ridesourcing platform (Figure \ref{fig:reg-costs-six}a). The markets corresponding to these scenarios more quickly reach a state in which the number of new registrations is negligible in terms of its effect on the daily number of participating drivers.

We observe that the marginal decrease in registration volume when $C_d$ grows is especially large when registration costs are limited. In a scenario without registration costs (scenario \textit{RC0}), nearly 900 drivers register with the platform, compared to approximately 430 when registration costs add up to \euro20 per day, and just over 200 when the daily registration penalty amounts to \euro40.  The consequence is that registration costs lead to reduced participation (Figure \ref{fig:reg-costs-six}b) and ultimately to a higher average experienced (Figure \ref{fig:reg-costs-six}c) and expected (Figure \ref{fig:reg-costs-six}d) income. Registration costs can thus be a crucial factor for whether drivers, on average, end up earning above or below the reservation wage. However, considering that registration costs need to be subtracted from the income of drivers, a scenario with $C_d$ equal to 40 still turns out to be least favourable for drivers, as demonstrated by Figure \ref{fig:reg-costs-six}e. In this scenario drivers that participate earn back on average 75\% of their total costs (including the cost of participation and registration), compared to 88\% when registration does not bear any costs and the total costs are made up of the reservation wage (and operational costs). This, however, considers only the income of participating drivers. It should not be forgotten that also registered drivers that do not participate on a given day end up with a negative daily profit due to their capital registration costs. In case they cannot easily discard their registration costs, for example by selling their car, it might still be their best option to keep participating, even when this results in a negative net income. Due to reduced supply, travellers may also be worse off when a ridesourcing service comes with high registration barriers for drivers (Figure \ref{fig:reg-costs-six}d), the extent to which likely dependent on context-specific variables. In our particular experiment, travel times are hardly affected by registration costs.

The results imply that ridesourcing providers, drivers and travellers may also suffer from high entry barriers for potential suppliers. Consequently, policies that aim at reducing the costs related to registration may be beneficial, for example offering affordable vehicle insurance deals to drivers.

\subsection{System optimum supply and user equilibrium solutions}
\begin{figure}[] 
  \centering
  \includegraphics[width=\textwidth]{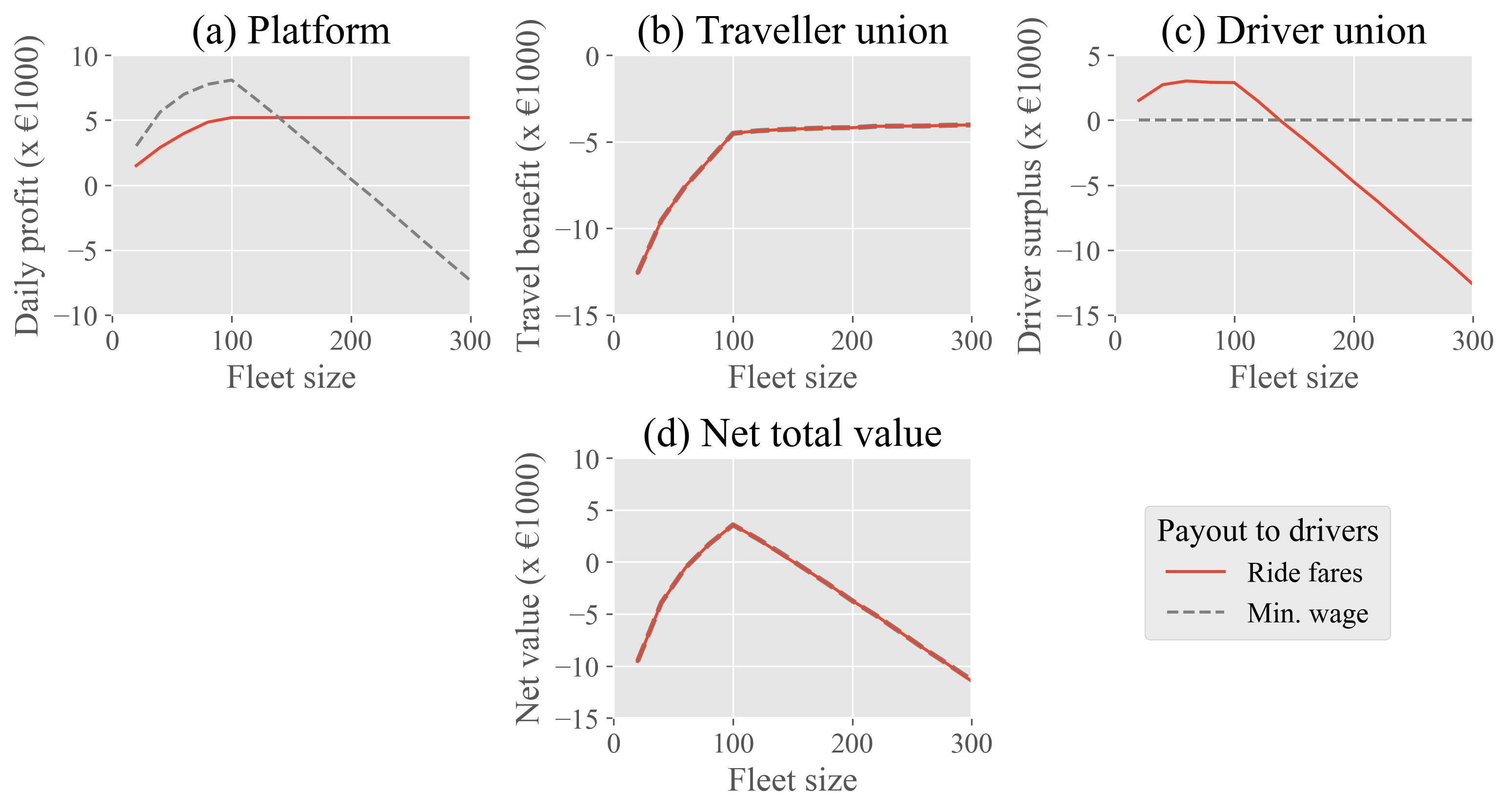}
  \caption{Optimality of supply in a system with fare-based driver payouts (assuming a platform commission rate of 25\%) and a system with wage-based driver payouts, for (a) the service provider, aiming at maximum profit, (b) the traveller union, minimising costs from waiting and rejected requests, (c) the driver union, maximising driver earnings over the reservation wage, and (d) an authority that evaluates the three previous objectives equally, maximising the summed net value}\label{fig:soue-four}
\end{figure}

In this section, we elaborate on the social optimality of a decentralised ridesourcing supply and discuss the implications for how regulation should be designed to safeguard the interests of different stakeholders in the process. The user-equilibrium solution obtained from our model is compared with the system optimum supply-level that is obtained from a brute force search for the optimal fleet size. Figure \ref{fig:soue-four}a shows the profit of a ridesourcing platform for different participation levels. Next to the typical ridesourcing scenario in which self-employed drivers get paid based on the rides they serve, we consider an alternative scenario in which drivers, instead, earn a guaranteed hourly wage, while also getting their operational costs reimbursed. Comparing platform profit in both scenarios, we observe a major difference in the financial consequences of oversupply for the service provider. In the typical ridesourcing scenario with fare-based payouts, oversupply does not induce additional costs, because ridesourcing providers pay drivers based on served demand, not participation. In the event of abrupt market contraction (e.g. pandemic crisis), for example, compared to service providers with employed drivers, ridesourcing providers benefit from reduced driver payouts that will partially offset the lower earnings from fares. Hence, a consequence of transaction-based driver payments is that, in contrast to more traditional transit providers paying drivers based on the number of hours worked, ridesourcing providers lack an incentive to curb their supply. In fact, as Figure \ref{fig:soue-four}b shows, they can benefit from oversupply as it leads to lower travel times for travellers, and thus, potentially, increased demand. These benefits are, however, relatively limited after supply reaches a specific point, which appears to be the minimum supply for which (nearly) all requests can be served. More supply will result in more efficient matches between drivers and travellers, yet yielding a minor effect on the travel times for riders in the system. 

Figure \ref{fig:soue-four}a also shows that fare-based driver payments are not necessarily optimal for the service provider. In the presented scenario, the service provider would actually be better off paying an hourly wage to a relatively limited number of drivers, thereby earning the full share of ride fares, than allowing self-employed drivers to collect these fares in return for a fee. It should be noted that this particular example does not consider that employed drivers may be entitled to social benefits and that drivers may not be willing to work for the minimum wage.

When taking the driver perspective, we find that the optimal fleet size in the fare-based scenario is relatively low (Figure \ref{fig:soue-four}c), peaking between 40 and 100 participating drivers. If supply is even lower, a lot of potential income is lost due to rejected requests, however, if it is higher, excessive competition leads to incomes below the reservation wage, and consequently, dissatisfied drivers. For supply volumes over 120 the total driver surplus is in fact negative. Yet, remarkably, in the reference scenarios of our experiment with decentralised supply, we find average daily participation volumes of approximately 150 drivers in equilibrium. This demonstrates that in the ridesourcing market the notion of 'the tragedy of the commons' may apply, in which the self-interested labour decisions of individual drivers lead to a suboptimal result for the whole group: excessive competition for rides and ultimately low payouts.

If we consider a society in which the societal value of a single monetary unit is independent of the party that it is assigned to, i.e. an extra profit of one euro for the platform or a single driver has the same value as a travel cost saving of one euro for one traveller, we find that the optimal ridesourcing fleet size for our particular experiment is 100 drivers, as illustrated by the total net value sketched in Figure \ref{fig:soue-four}d. Lower supply levels are undesired from the platform's and travellers' perspective, while higher supply leads to a significantly deteriorated driver income with only a very limited benefit for travellers. The social optimum in this case is thus considerably lower than the user equilibrium, which depicts the potential value of supply caps in ridesourcing markets. Although ridesourcing providers are typically reluctant to accept the implementation of supply caps, our analysis illustrates that their negative effect on rider level of service and ultimately platform profit may be very limited, especially in a saturated market. In this particular case, a reduction of supply from 300 to 100 drivers only induces a single minute of extra waiting time per request. 

We note that the socially optimal fleet size for ridesourcing services is equal to that of a transit service with employed drivers, because the objective function for the net total value ultimately contains the same elements: revenue from fares, operational costs and labour participation costs. The only difference is the distribution of those over different stakeholders. If a society indeed considers a single monetary unit equally valuable to all stakeholders, it can thus be stated that only the fleet size of an on-demand transit service matters from a societal perspective, not whether drivers are paid for participating or based on the travel requests they satisfied.

\subsection{Model sensitivity}
\subsubsection{Learning}
Learning parameter $\omega$ indicates how drivers value recent experiences compared to preceding experiences over time. A low value of $\omega$ corresponds to a situation in which drivers assign a relatively high value to their recent experience (see Equation \ref{eq:learning-2}), for example because they believe old experiences are not representative for the present state of the system or because they cannot perfectly memorize their income from previous days. In contrast, if $\omega$ goes to infinity, drivers' expected income equals their average experienced income. In this study, we assumed $\omega$ to be equal to 5, indicating that the weight of new experiences decreases to 0.2 within 5 days, and stays constant thereafter. To establish to what extent the results presented in this section are specific to the learning parameter, we have repeated the experiment for the reference scenarios, while varying the value of the learning parameter $\omega$ between 3 and 100.

\begin{figure}[] 
  \centering
  \includegraphics[width=\textwidth]{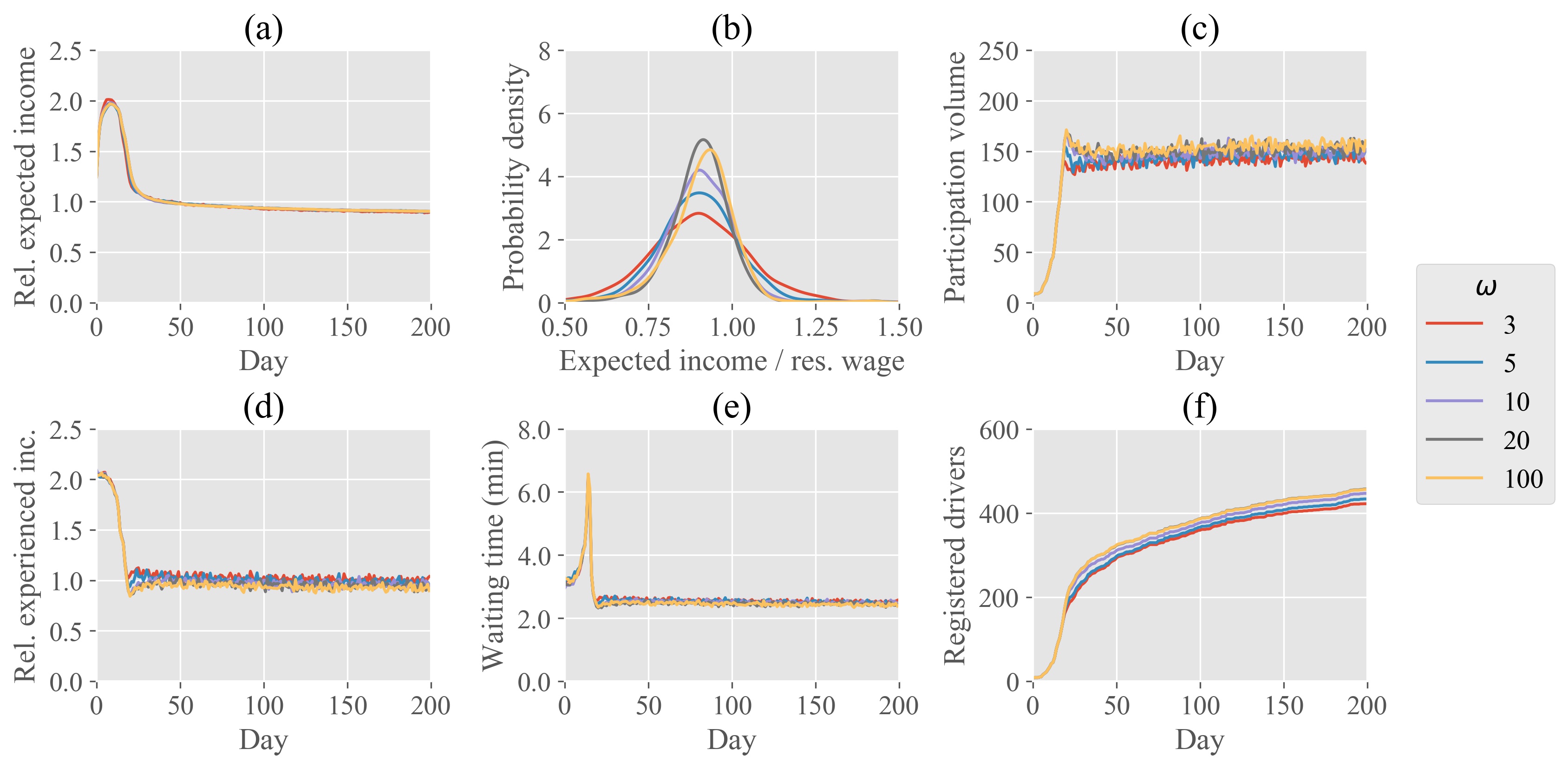}
  \caption{The effect of the rate of learning on (a) the evolution of the average expected income of registered drivers as ratio of their reservation wage, (b) the probability density function of expected income (as ratio of their reservation wage) for registered drivers, (c) the evolution of daily participation volumes, (d) the evolution of average experienced income of participating drivers as ratio of their reservation wage, (e) the evolution of the average waiting time for pick-up for travellers and (f) the evolution of the total number of registered drivers}\label{fig:omega-six}
\end{figure}

We find that $\omega$ has a limited effect on ridesourcing provision. One of the notable differences is that, although the mean expected income in equilibrium is unaffected by $\omega$ (Figure \ref{fig:omega-six}a), the distribution of expected income over registered drivers differs (Figure \ref{fig:omega-six}b). This can be explained by the fact that when $\omega$ is small, drivers are more likely to 'overreact' to a single negative experience, resulting in a pool of 'unsatisfied' drivers with expected income levels significantly below the reservation wage. These drivers will not be tempted to participate again, limiting participation on the platform (Figure \ref{fig:omega-six}c) and driving up the experienced income (Figure \ref{fig:omega-six}d) and ultimately the expected income of the other registered drivers (Figure \ref{fig:omega-six}b). It also results in a minor difference in the average waiting time for travellers (Figure \ref{fig:omega-six}e). Moreover, we establish that registration volumes slightly diverge in an early stage of adoption (Figure \ref{fig:omega-six}f). The reason is that when $\omega$ is small, drivers more quickly observe that earnings are dropping (Figure \ref{fig:omega-six}d), which they communicate to drivers that have not yet registered. Nevertheless, the effect of $\omega$ was found to be limited and we do not expect a major impact on the main findings regarding dynamics in ridesourcing supply.

\subsubsection{Information diffusion}
\begin{figure}[] 
  \centering
  \includegraphics[width=\textwidth]{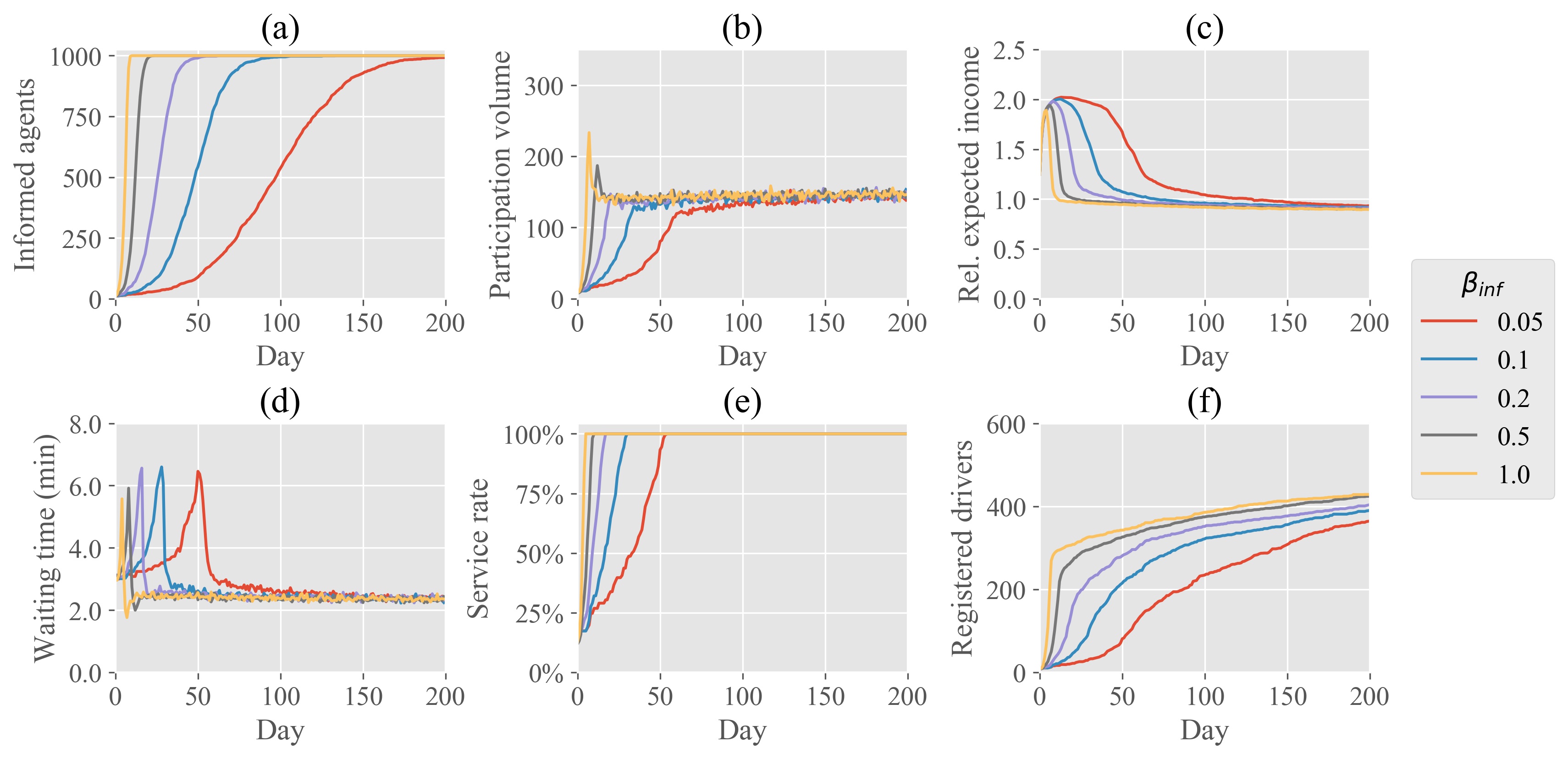}
  \caption{The effect of the information transmission rate on the evolution of (a) the number of informed agents, (b) daily participation volumes, (c) the average expected income of registered drivers as ratio of their reservation wage, (d) the average waiting time for pick-up for travellers, (e) platform profit and (f) the total number of registered drivers}\label{fig:inf-beta-six}
\end{figure}

This study considers that drivers need to become aware about the existence of a ridesourcing service before they can supply labour to it. To this end, we introduce an information diffusion process with a transmission rate $\beta_{\mathrm{inf}}$ of 0.2. Lacking empirical evidence of the specification of the information diffusion process, we need to test whether our findings also apply under different diffusion settings. We test four alternative values for $\beta_{\mathrm{inf}}$, ranging between 0.05 and 1.0. We find that, given a value for $\beta_{\mathrm{inf}}$ that allows (nearly) all drivers to be informed at the end of the simulation (Figure \ref{fig:inf-beta-six}a), the specification of the diffusion process has hardly any effect on labour supply in equilibrium. The different scenarios for $\beta_{\mathrm{inf}}$ converge to the same participation volume (Figure \ref{fig:inf-beta-six}b), with a similar average expected income (Figure \ref{fig:inf-beta-six}c), average waiting time for travellers (Figure \ref{fig:inf-beta-six}d) and service rate (Figure \ref{fig:inf-beta-six}e), which demonstrates the generalisability of the results, concerning the value of $\beta_{\mathrm{inf}}$. Although the indicators are similar in equilibrium, we note clear differences in the adoption process. When $\beta_{\mathrm{inf}}$ is high, many drivers become aware about the service at the same time. In an early phase of adoption (phase 1 as introduced in Subsection \ref{subsect:phases}), when there are few drivers supplying labour to the platform and income levels are high, this leads to a big registration peak (Figure \ref{fig:inf-beta-six}f) and excessive participation, with relatively low driver incomes and limited waiting times for travellers. In scenarios with a lower information transmission rate we do not observe such a peak in participation, but rather a steady increase towards the equilibrium value. As a consequence, in scenarios in which communication about innovations takes place slowly, it takes longer before the level of service reaches a satisfactory level, with the large majority of rides accepted and a relatively low average waiting time for travellers.

\section{Conclusions}
\subsection{Study significance}
This study is pioneering in analysing the dynamics of (decentralised) ridesourcing supply while accounting for labour supply decisions considering both long-term platform registration and short-term participation. Our platform registration submodel considers that registration requires information about earnings, and that it comes with one-off registration costs like insurance and vehicle acquisition, which are sunk in subsequent participation decisions. With a probabilistic participation choice submodel, we account for unobserved variables in the decision to work on a given day, like planned activities for this particular day. The model is applied to the case of Amsterdam in order to investigate the effect of supply market properties, platform pricing and supply-side entry barriers on the evolution of ridesourcing supply. In addition, we comment on the optimality of decentralised ridesourcing supply from the perspectives of drivers, travellers and service provider, based on an exhaustive search. 

The results demonstrate that labour supply in ridesourcing may be non-linear and undergo several transitions, hereby inducing significant variations in average income, profit and level of service. It highlights the need for models capturing dynamic interactions in ridesourcing provision, such as the one presented in this work. 

\subsection{Key findings}
\paragraph*{Fleet size.}
We find that in a decentralised system, as long as drivers earn a competitive income and not yet all potential drivers are registered, new suppliers are attracted to the market at a relatively high pace. For the base scenario of our experiment, this phenomenon results in an equilibrium participation volume of 150 drivers. With this level of supply, there is relatively strong competition for pick-ups, resulting in payouts below drivers' reservation wages. Instead, for the community of (potential) ridesourcing drivers in our experiment, a fleet size of 40 - 100 drivers is considered to be optimal. Such a solution implies that the fewer drivers participating will earn a significantly higher income. The above findings demonstrate that the tragedy of the commons may apply in ridesourcing provision, in which the self-centered labour decisions of individuals ultimately harm the common interests of the group.

Unlike traditional transit providers with employed drivers, ridesourcing providers lack a direct financial incentive to curb supply. Our results demonstrate however that there may be an alternative balancing loop in ridesourcing supply, i.e. profit-maximising service providers may be best off claiming a relatively high rate on fares collected through their platforms, even when this means that fewer drives will participate and, consequently, that a portion of the travel requests has to be rejected. In our experiment, in equilibrium approximately 60 drivers participate when a platform opts for a profit-maximising commission rate of 45\%, compared to 180 drivers when the commission rate is 25\%. This results in a decline in the probability that a request can be matched from 100\% to 85\%, and in an increase in the average waiting time from 2 to 6 minutes. Remarkably, average drivers earnings in the experiment are hardly affected by the commission rate of the platform. The rationale here is that the influx of new drivers on the platform is limited when the commission rate is high. This implies that registration barriers may mitigate the tragedy of the commons in ridesourcing supply.

\paragraph*{Labour market effect.}
The expected income is especially low when the average reservation wage is low. In this case, drivers are relatively quick to register, leading to a fierce competition and ultimately a decreasing income for those already registered. Free-lance workers in the market will thus suffer from a shrinking economy in which other labour opportunities are scarce. The same applies to a labour market with large inequalities, in which ridesourcing services are flooded with drivers that have limited opportunities in the market, and are willing to work even when earnings are low.

\subsection{Policy implications}
\paragraph*{Supply regulation.} 
Similarly to the results of the semi-dynamic model by \cite{yu2020balancing}, our findings provide support for the potential effectiveness of a supply cap, which has for example been implemented in New York City. It may push earnings over the reservation wage without significantly impeding travellers' waiting times. At the same time, our results show that the value to which the cap is set is crucial. For instance, in our experiment, supply caps above 400 drivers or more would yield no effect on driver income. On the other hand, we find that when supply caps are too restrictive, they may be detrimental to the level of service offered by the platform. This is in line with the results of the queuing theoretic equilibrium model formulated by \cite{li2019regulating}, demonstrating that a supply cap can lead to reduced driver earnings when too many consumers leave the market. In any case, given that capital registration costs jeopardise the income of drivers, transit authorities should avoid supply caps that assign an additional cost to operation under the supply cap.

\paragraph*{Pricing strategy.}
A profit-maximising platform will increase its commission rate up to the point that so many drivers opt out that lost commission from rejected requests outweighs the higher revenue on remaining requests. Our experiments demonstrate that at this point already a significant portion of ride requests may need to be rejected. In addition, we find that such a profit-maximising strategy may result in relatively long waiting times for travellers. These results suggest that the pricing strategy of a ridesourcing platform may need to be regulated. Our results in fact demonstrate that this may be highly beneficial from a societal standpoint, given that a near-optimal profit can be achieved with a significantly lower commission rate, yielding a much improved level of service for travellers. This confirms earlier findings based on an analytical economic model by \cite{zha2016economic} regarding the effectiveness of regulation of the commission in increasing the social welfare generated by ridesourcing platforms.

\subsection{Future research}
In this study, we focus on supply evolution in order to understand its dynamics and describe emerging phenomena, which can be further embedded in models of co-evolution of supply and demand. An interesting direction for future research is the extent to which outcomes of a monopolistic market are also applicable to markets in which service providers compete for supply and demand. For example, future research may consider how supply evolution is affected by aggressive penetration pricing strategies aimed at pushing other service providers out of the market. It may also be interesting to analyse how external shocks to the market lead to swings in the transition process. Our model can be extended to study supply evolution of ridesourcing services offering pooled rides, which will affect the income of participating drivers. In essence, our approach with a day-to-day shell and a core capturing within-day dynamics allows to analyse ridesourcing supply evolution under various operational within-day strategies.

As a concluding remark, we stress the need for more empirical evidence on labour supply by ridesourcing drivers, as model input - based on cross-sectional data - and for validation of the results - based on longitudinal data. Enhancing the empirical underpinning on labour supply behaviour by (potential) ridesourcing drivers will support the specification of a simulation framework like the one presented here and thereby allow to significantly improve our knowledge on ridesourcing implications for drivers, travellers, platforms and society at large.


\section*{Acknowledgement(s)}
Preliminary versions of this paper have been accepted for presentation at the 2020 hEART conference in Lyon and the 2021 TRB Annual Meeting in Washington D.C.

\section*{Funding}
This work was supported by the European Research Council under Grant 804469 and by the Amsterdam Institute for Advanced Metropolitan Solutions.

\bibliographystyle{tfcad}
\bibliography{interactcadsample}

\end{document}